\documentclass[aps,prd,superscriptaddress,10pt,showpacs,notitlepage,  
nofootinbib
]{revtex4-1}
\usepackage{bm}
\usepackage{amsmath,amssymb,amsthm}
\usepackage{latexsym,graphicx,color,subfigure}
\usepackage{enumerate}
\usepackage{amssymb}
\usepackage{hyperref}
\usepackage{mathtools}

\usepackage{graphicx}
\graphicspath{ {images/} }

\usepackage{color}
\usepackage{soul}

\newcommand{\be}{\begin{equation}}
\newcommand{\ee}{\end{equation}} 
\newcommand{\beq}{\begin{eqnarray}}
\newcommand{\eeq}{\end{eqnarray}}

\newcommand{\bea}{\begin{eqnarray}}
\newcommand{\eea}{\end{eqnarray}}

\renewcommand{\vec}[1]{\boldsymbol{#1}}

\allowdisplaybreaks[3]
\def\simge{\mathrel{
   \rlap{\raise 0.511ex \hbox{$>$}}{\lower 0.511ex \hbox{$\sim$}}}}
\def\simle{\mathrel{
   \rlap{\raise 0.511ex \hbox{$<$}}{\lower 0.511ex \hbox{$\sim$}}}}
\def\bigs{\mathrel{
   \rlap{\raise 0.531ex \hbox{$>$}}{\lower 0.531ex \hbox{$<$}}}}

\usepackage[normalem]{ulem}

\renewcommand\sout{\bgroup \color{red} \ULdepth=-.5ex \ULset}
\renewcommand\st{\bgroup \color[rgb]{0.117, 0.564, 1.000} \ULdepth=-.5ex \ULset}

\newcommand{\drm}{\mathrm{d}}

\usepackage{framed}
\usepackage{bigints}
\usepackage{ulem}

\begin{document} 
\title{Spin-isospin Kondo effects for $\Sigma_{c}$ and $\Sigma_{c}^{\ast}$ baryons and $\bar{D}$ and $\bar{D}^{\ast}$ mesons}
\author{Shigehiro Yasui}
\email{yasuis@keio.jp}
\affiliation{Research and Education Center for Natural Sciences,\\ Keio University, Hiyoshi 4-1-1, Yokohama, Kanagawa 223-8521, Japan}
\author{Tomokazu Miyamoto}
\email{tomokazu.miyamoto10@physics.org, tomokazu.miyamoto2@keio.jp}
\affiliation{Research and Education Center for Natural Sciences,\\ Keio University, Hiyoshi 4-1-1, Yokohama, Kanagawa 223-8521, Japan}
\affiliation{Graduate School of Engineering Science, Yokohama National University, Tokiwadai 79-1, Yokohama, Kanagawa 240-8501, Japan}
\date{\today}
\begin{abstract}
We study the Kondo effect for a $\Sigma_{c}$ ($\Sigma_{c}^{\ast}$) baryon in nuclear matter.
In terms of the spin and isospin ($\mathrm{SU}(2)_{\mathrm{spin}} \times \mathrm{SU}(2)_{\mathrm{isospin}}$) symmetry, the heavy-quark spin symmetry and the S-wave interaction, we provide the general form of the Lagrangian for a $\Sigma_{c}$ ($\Sigma_{c}^{\ast}$) baryon and a nucleon.
We analyze the renormalization equation at the one-loop level, and find that the coexistence of spin exchange and isospin exchange magnifies the Kondo effect in comparison with the case where the spin-exchange interaction and the isospin-exchange interaction exist separately.
We demonstrate that the solution exists for the ideal sets of the coupling constants, including the $\mathrm{SU}(4)$ symmetry as an extension of the $\mathrm{SU}(2)_{\mathrm{spin}} \times \mathrm{SU}(2)_{\mathrm{isospin}}$ symmetry.
We also conduct a similar analysis for the Kondo effect of a $\bar{D}$ ($\bar{D}^{\ast}$) meson in nuclear matter.
On the basis of the obtained result, we conjecture that there could exist a ``mapping" from the heavy meson (baryon) in vacuum onto the heavy baryon (meson) in nuclear matter.
\end{abstract}
\maketitle


\setlength\arraycolsep{2pt}

\section{Introduction}

In 1964, J.~Kondo explained why the electrical resistance in the metal which contains some impurity atoms with a nonzero spin increases logarithmically at low temperatures~\cite{Kondo:1964}.
The logarithmic increase of the electrical resistance with the heavy impurity occurs when the following conditions are satisfied: (i) Fermi surface (degenerate state), (ii) particle-hole creation (loop effect), and (iii) non-Abelian interaction (e.g. the spin-exchange interaction)~\cite{Hewson,Yosida,Yamada}.
It is understood that under these three conditions, the coupling constant for the interaction becomes stronger, and the Landau pole appears.
Since his work was recognized, the Kondo effect has had wider implications for theoretical approaches in quantum systems: the renormalization group method~\cite{Anderson1970}, the numerical renormalization group~\cite{Wilson:1974mb}, the Bethe ansatz~\cite{PhysRevLett.45.379,Weigmann,RevModPhys.55.331}, the boundary conformal field theory~\cite{Affleck:1990zd,Affleck:1991tk,Affleck:1990by,Affleck:1990iv,Affleck:1992ng,Ludwig:1994nf,Affleck:1995ge}, the bosonization method~\cite{FRADKIN1989710,Ye:1996dj,PhysRevB.9.2911,PhysRevLett.81.196,PhysRevB.61.6918}, the mean-field approximation (the large $N$ limit)~\cite{Takano:1966,Yoshimori:1970,Lacroix:1979,PhysRevB.28.5255,ReadNewns1983,PhysRevLett.57.877,PhysRevB.35.3394,PhysRevB.35.5072,PhysRevLett.79.4665,PhysRevB.58.3794,PhysRevLett.85.1048,Eto:2001,PhysRevB.75.132407,Yanagisawa:2015conf,Yanagisawa:2015}, and so on.

The Kondo effect is not simply studied in condensed matter physics, but is applicable to the nuclear physics where the strong interaction plays a role of the main fundamental force.\footnote{At an early stage, the Kondo effect was studied for deformed nuclei, where the itinerant fermion is a nucleon and the impurity is played by the deformed nucleus~\cite{Sugawara-Tanabe:1979}. The non-Abelian interaction is provided by the spin exchange through the Coriolis force. However, it leads to the suppression of the Kondo effect due to the sign of the coupling constant.}
For example, we consider the case where the heavy hadrons involving charm or bottom flavor are brought into the nuclear matter (see Refs.~\cite{Hosaka:2016ypm,Krein:2017usp} for reviews).
They can be regarded as the heavy impurity particles, because their masses are much heavier than the light (up, down, and strange) quarks.
Several heavy hadrons have been used in the previous studies: a $\bar{D}$ meson ($D^{-}$ or $\bar{D}^{0}$ meson) and a $\bar{D}^{\ast}$ meson ($D^{\ast-}$ or $\bar{D}^{\ast0}$ meson)~\cite{Yasui:2013xr,Yasui:2016ngy}, or a $D_{s}^{-}$ meson and a $D_{s}^{\ast -}$ meson~\cite{Yasui:2016hlz}, in charm flavor.
It is certainly true that the heavy hadrons are not stable, because they can decay into the light hadrons via the weak interaction.
Nevertheless, it is worth considering the heavy hadrons in the nuclear matter when we only consider the strong interaction or the electromagnetic interaction.
The heavy hadrons may be produced in atomic nuclei experimentally at the high-energy accelerator facilities.
Clearly, the conditions (i) and (ii) for the Kondo effect are met; the Fermi surface and the particle-hole creations exist in the nuclear matter at the low temperatures.
When it comes to the condition (iii), the non-Abelian interaction is provided by the spin-exchange interaction and/or by the isospin-exchange interaction, both of which obey the $\mathrm{SU}(2)_{\mathrm{spin}}$ symmetry and/or the $\mathrm{SU}(2)_{\mathrm{isospin}}$ symmetry, respectively.
The research on the Kondo effect for the $\bar{D}$ and $\bar{D}^{\ast}$ mesons and the $D_{s}^{-}$ and $D_{s}^{\ast -}$ mesons in nuclear matter was conducted by using the perturbative calculation~\cite{Yasui:2013xr} and the mean-field approximation~\cite{Yasui:2016hlz}.
The Kondo effect for the heavy hadron in atomic nuclei was studied in terms of the mean-field approximation in the Lipkin model, in which the fluctuation effect was also considered~\cite{Yasui:2016ngy}.

From a QCD perspective, it is noteworthy that the Kondo effect was also studied for a charm or bottom quark in quark matter, where the non-Abelian interaction between the heavy quark and the itinerant light quark is provided by the color-exchange interaction in accordance with the $\mathrm{SU}(3)_{\mathrm{color}}$ symmetry~\cite{Yasui:2013xr,Hattori:2015hka}.
This is called the QCD Kondo effect~\cite{Hattori:2015hka}.
The QCD Kondo effect was studied in various theoretical methods: the simple perturbation~\cite{Yasui:2013xr}, the (perturbative) renormalization group with gluon exchange~\cite{Hattori:2015hka}, the mean-field approximation~\cite{Yasui:2016svc,Yasui:2016yet,Yasui:2017izi,Yasui:2017bey}, the conformal boundary theory~\cite{Kimura:2016zyv,Kimura:2018vxj}.
The competition between the QCD Kondo effect and the color superconductivity or the chiral condensate was analyzed~\cite{Kanazawa:2016ihl,Suzuki:2017gde}.
In addition, the transport properties such as the electric conductivity and the shear viscosity were studied~\cite{Yasui:2017bey}.
It is important to mention that the QCD Kondo effect in the quark matter with the light flavor $N_{f}\ge2$ serves the overscreened Kondo effect instead of the normal Kondo effect with an exact screening, and it leads to the non-Fermi liquid behavior~\cite{Kanazawa:2016ihl,Kimura:2016zyv,Kimura:2018vxj}.
The heavy quark in strong magnetic field induces the QCD Kondo effect at the vanishing density (the magnetically induced QCD Kondo effect), where the light quarks are confined with degeneracy in the lowest Landau level~\cite{Ozaki:2015sya}.
It was recently argued that the QCD Kondo effect occurs even in the absence of the heavy quark in quark matter: the color nonsinglet gapped quark in the two-flavor superconductivity (2SC) plays the role of the ``heavy impurity", and it leads to the non-Abelian interaction with the light ungapped quarks which do not participate to form the 2SC gap~\cite{Hattori:2019zig}.

The purpose of the present paper is to study the Kondo effect for a $\Sigma_{c}$ ($\Sigma_{c}^{\ast}$) baryon in nuclear matter.
The $\Sigma_{c}$ ($\Sigma_{c}^{\ast}$) baryon has spin 1/2 (3/2) and isospin 1, and it can provide the non-Abelian interaction by the spin and isospin-exchange with a nucleon.
We consider the heavy mass limit for the heavy quark (a charm quark)~\cite{Isgur:1989vq,Isgur:1989ed,Isgur:1991wq}, where the spin-flip and isospin-flip interactions work on the light component in the $\Sigma_{c}$ ($\Sigma_{c}^{\ast}$) baryon, i.e., the light diquark ($qq$).
Indeed, the spin-flip process for the heavy quark is suppressed by the factor $\Lambda_{\mathrm{QCD}}/m_{Q}$ with $\Lambda_{\mathrm{QCD}}$ being the low-energy scale of the QCD and $m_{Q}$ being the mass of the heavy quark.
Thus, the spin of a heavy quark can be regarded as the conserved quantity in the heavy-quark mass limit.
This is called the heavy-quark spin (HQS) symmetry~\cite{Isgur:1989vq,Isgur:1989ed,Isgur:1991wq} (see also Refs.~\cite{Manohar:2000dt}).
In the present study, we consider only the leading-order term in the heavy-quark mass limit, and neglect the corrections at ${\cal O}(\Lambda_{\mathrm{QCD}}/m_{Q})$.
For example, the heavy quark symmetry is seen approximately in the small mass splitting between a $\Sigma_{c}$ baryon and a $\Sigma_{c}^{\ast}$ baryon (about 65 MeV) which is much smaller than the baryon masses (2286 and 2520 MeV).
The HQS will provide us with a good approximation as the first step to investigate the Kondo effect for the $\Sigma_{c}$ ($\Sigma_{c}^{\ast}$) baryon.
The effective theory of the $\Sigma_{c}$ ($\Sigma_{c}^{\ast}$) baryon can be constructed in a general form when we follow the HQS symmetry~\cite{Falk:1990yz,Falk:1991nq,Burdman:1992gh,Wise:1992hn,Yan:1992gz,Cho:1992gg} (see also Refs.~\cite{Manohar:2000dt,Casalbuoni:1996pg} for reviews), and this formalism will be applied to the interaction between a $\Sigma_{c}$ ($\Sigma_{c}^{\ast}$) baryon and a nucleon.
Given the fact that the $\Sigma_{c}$ ($\Sigma_{c}^{\ast}$) baryon has two different non-Abelian interactions of spin and isospin, i.e., the $\mathrm{SU}(2)_{\mathrm{spin}} \times \mathrm{SU}(2)_{\mathrm{isospin}}$ symmetry, we will see that those two symmetries induce rich structures of the Kondo effect.
As an ideal situation, for example, the $\mathrm{SU}(2)_{\mathrm{spin}} \times \mathrm{SU}(2)_{\mathrm{isospin}}$ symmetry will provide the $\mathrm{SU}(4)$ symmetry by tuning the coupling constants in the interaction term appropriately.
Throughout the present study, we will perform the analysis based on the renormalization group (RG) equation, namely the poor man's scaling method, as the simple perturbative method~\cite{Anderson1970}.
The Kondo effect induces an enhancement of the coupling constant at around the low-energy scale, known as the Kondo scale, for the $\Sigma_{c}$ or $\Sigma_{c}^{\ast}$ baryon (the $\bar{D}$ or $\bar{D}^{\ast}$ meson) in nuclear matter. Our main goal is to indicate the existence of the Kondo scale.
The observables which are relevant to the Kondo effect could be the transport coefficients, such as the heat conductivity, the electrical resistance, and the shear viscosity, because the enhanced coupling strength can affect these quantities drastically at around the Kondo scale.
Other possible observables are the change of the nuclear structure in atomic nuclei: the modifications of  the excitation spectra stemming from the enhanced coupling between a nucleon and a $\Sigma_{c}$ or $\Sigma_{c}^{\ast}$ baryon (a $\bar{D}$ or $\bar{D}^{\ast}$ meson). 
These observables are related to the dynamical and static properties of the system which is geared to the Kondo effect.

Several comments are in order.
In the literature, the binding of a $\Sigma_{c}$ ($\Sigma_{c}^{\ast}$) baryon in nuclear matter was estimated by the QCD sum rules~\cite{Wang:2011yj,Azizi:2018dtb}. The present discussion about the Kondo effect will be useful for further investigation on the binding energy.
We notice that a $\Lambda_{c}$ baryon is not relevant to the Kondo effect in contrast to the $\Sigma_{c}$ ($\Sigma_{c}^{\ast}$) baryon, because the light diquark ($qq$) in the $\Lambda_{c}$ baryon has spin 0 and isospin 0, and there is no exchange interaction of spin and isospin between the baryon and a nucleon, as it was analyzed in Ref.~\cite{Yasui:2018sxz} (see also the recent work~\cite{Carames:2018xek,Vidana:2019amb}).\footnote{Those studies rely on the $\Lambda_{c}N$ interaction strength estimated by the lattice QCD simulations~\cite{Miyamoto:2017tjs} and the chiral extrapolations~\cite{Haidenbauer:2017dua}. The obtained binding energy for a $\Lambda_{c}$ baryon is consistent with the results by the QCD sum rules~\cite{Ohtani:2017wdc}.}
Bottom hadrons, which are in general heavier than charm hadrons, could be more suitable for studying the Kondo effect; however, we will not repeat the same discussion for the bottom hadrons.
Replacing a $\Sigma_{c}$ ($\Sigma_{c}^{\ast}$) baryon by a $\Sigma_{b}$ ($\Sigma_{b}^{\ast} $) baryon is a straightforward task, although it would provide more favorable conditions for greater accuracy of the HQS symmetry.
The greater accuracy is seen directly in the mass splitting between a $\Sigma_{b}$ baryon and a $\Sigma_{b}^{\ast}$ baryon (about 20 MeV) in comparison to their masses (5810 MeV and 5830 MeV, respectively).

The paper is organized as follows.
In Sec.~\ref{sec:Lagrangian}, we introduce the Lagrangian for a $\Sigma_{c}$ ($\Sigma_{c}^{\ast}$) baryon and a nucleon.
We suppose the $\mathrm{SU}(2)_{\mathrm{spin}} \times \mathrm{SU}(2)_{\mathrm{isospin}}$ symmetry, the HQS symmetry, and the S-wave interaction.
In Sec.~\ref{sec:RG_eq_Sigmac}, 
we carefully investigate the solutions of the RG equation,
 and point out that the simultaneous flipping of the spin and the isospin is important for magnifying the Kondo effect.
In Sec.~\ref{sec:revising_D}, we revisit the Kondo effect for a $\bar{D}$ ($\bar{D}^{\ast}$) meson in nuclear matter, where the similar analysis is applicable.
In Sec.~\ref{sec:K_mapping}, we surmise that the Kondo effect induces a mapping between the heavy meson (baryon) in vacuum and the heavy baryon (meson) in nuclear matter.
The final section is devoted to the conclusion.

\section{Lagrangian for a $\Sigma_{c}$ ($\Sigma_{c}^{\ast}$) baryon and a nucleon}
\label{sec:Lagrangian}

We begin by considering the nuclear matter in which a $\Sigma_{c}$ ($\Sigma_{c}^{\ast}$) baryon exists as an impurity particle, assuming that the nuclear matter is approximately regarded as the free fermion gas where the nucleon is described by the nonrelativistic two-component spinor field $\varphi(x)$.
We follow the procedures for the construction of the field of the heavy hadron based on the HQS symmetry~\cite{Falk:1990yz,Falk:1991nq,Burdman:1992gh,Wise:1992hn,Yan:1992gz,Cho:1992gg} (see also Refs.~\cite{Manohar:2000dt,Casalbuoni:1996pg} for reviews), and apply this formalism to the interaction between a $\Sigma_{c}$ ($\Sigma_{c}^{\ast}$) baryon and a nucleon.
In this framework, the field of the $\Sigma_{c}$ ($\Sigma_{c}^{\ast}$) baryon can be decomposed to the diquark part ($qq$) and the heavy quark part ($Q$), where the quantum number of the diquark is spin 1 and isospin 1.
We introduce the vector field $A^{\mu}(x)$ ($\mu=0,1,2,3$), which satisfies $v_{\mu}A^{\mu}=0$, for the diquark part.
We also introduce the effective heavy-quark field $u_{v}(x)$, which satisfies $v_{\nu}\gamma^{\nu}u_{v}=u_{v}$, for the heavy quark part.
We define $u_{v}(x)$ by $\displaystyle u_{v}(x)=\frac{1}{2}\bigl(1+\gamma_{\mu}v^{\mu}\bigr)e^{im_{Q}v\cdot x}u(x)$ in the $v$-frame with the four-velocity $v^{\mu}$ ($v^{0}>0$ and $v_{\mu}v^{\mu}$=1) and the heavy quark mass $m_{Q}$, where $u(x)$ is the original four-spinor heavy-quark field at $x$ in the four-dimensional coordinate system.
We consider that the sum is taken over the repeated indices.
The condition $v_{\nu}\gamma^{\nu}u_{v}=u_{v}$ stems from the requirement to project the field $u(x)$ to the positive-energy part.
It is supposed that the heavy quark is at rest in the coordinate frame with the four-velocity $v^{\mu}$.
In the following discussion, we consider the static frame by setting $v^{\mu}=(1,\vec{0})$.
With this setup, we define the composite field for the $\Sigma_{c}$ ($\Sigma_{c}^{\ast}$) baryon:
\begin{eqnarray}
   \Psi_{v}^{\mu}(x)=A^{\mu}(x)u_{v}(x).
\end{eqnarray}
Notice that $\Psi_{v}^{\mu}$ has only the off-mass-shell (residual) energy-momentum component with the energy scale smaller than the heavy-baryon mass, because the $\Sigma_{c}$ ($\Sigma_{c}^{\ast}$) baryon is supposed to be at rest in the $v$-frame.
We also notice that $\Psi_{v}^{\mu}$ satisfies $v_{\nu}\gamma^{\nu}\Psi_{v}^{\mu}=\Psi_{v}^{\mu}$ and $v_{\mu}\Psi_{v}^{\mu}=0$.
The former and latter properties are induced by $v_{\nu}\gamma^{\nu}u_{v}=u_{v}$ and $v_{\mu}A^{\mu}=0$, respectively.
With those two conditions, the number of degrees of freedom in $\Psi_{v}^{\mu}$ is $3\times2=6$.

In the above construction, $\Psi_{v}^{\mu}$ is a superposed state of the $\Sigma_{c}$ baryon (spin $1/2$) and the $\Sigma_{c}^{\ast}$ baryon (spin $3/2$).
This reflects the concept that the spin of the diquark and the spin of the heavy quark are good quantum numbers in the heavy-quark symmetry, and that the $\Sigma_{c}$ baryon and the $\Sigma_{c}^{\ast}$ baryon can be superposed.
In the physical space, it is convenient to introduce the fields of $\Sigma_{c}$ baryon and $\Sigma_{c}^{\ast}$ baryon by projecting $\Psi_{v}^{\mu}$ to the $\Sigma_{c}$ baryon component and the $\Sigma_{c}^{\ast}$ baryon component:
\begin{eqnarray}
   \Psi_{v1/2}=\frac{1}{\sqrt{3}} \gamma_{5} \gamma_{\mu} \Psi_{v}^{\mu},
\label{eq:Psi_2}
\end{eqnarray}
for the $\Sigma_{c}$ baryon and
\begin{eqnarray}
  \Psi_{v3/2}^{\mu} = \Psi_{v}^{\mu} - \frac{1}{3} \bigl( \gamma^{\mu} + v^{\mu} \bigr) \gamma_{\nu} \Psi_{v}^{\nu},
\label{eq:Psi_4}
\end{eqnarray}
for the $\Sigma_{c}^{\ast}$ baryon.
Equivalently, $\Psi_{v}^{\mu}$ is expressed as a sum of $\Psi_{v1/2}$ and $\Psi_{v3/2}$,
\begin{eqnarray}
   \Psi_{v}^{\mu} = \frac{1}{\sqrt{3}} \bigl( \gamma^{\mu} + v^{\mu} \bigr) \gamma_{5} \Psi_{v1/2} + \Psi_{v3/2}^{\mu}.
\end{eqnarray}
In the HQS formalism, the $\Sigma_{c}$ baryon and the $\Sigma_{c}^{\ast}$ baryon are degenerate in mass and are interchangeable to each other by the HQS symmetry.
For this reason, it is essential to consider a $\Sigma_{c}$ baryon and a $\Sigma_{c}^{\ast}$ baryon to be the effective degrees of freedom.
We will see that the heavy-quark-spin symmetry induces the mixing between the $\Sigma_{c}N$ state and the $\Sigma_{c}^{\ast}N$ state ($N$ for a nucleon) in the nuclear matter.

With the above setup, we consider the Lagrangian in the case where a nucleon and a $\Sigma_{c}$ ($\Sigma_{c}^{\ast}$) baryon interact with each other through the $S$-wave interaction on low-energy scale.
The $\Sigma_{c}N$ ($\Sigma_{c}^{\ast}N$) interaction was considered in the one-boson exchange model with a nonzero range~\cite{Maeda:2015hxa,Maeda:2018xcl}.
In contrast to them, we suppose that the $\Sigma_{c}N$ ($\Sigma_{c}^{\ast}N$) interaction is provided by the contact-type with a zero range.
The contact-type interaction and the HQS symmetry allow us to have the most general form of the Lagrangian:
\begin{eqnarray}
 {\cal L}[\varphi,\Psi_{v}^{i}]
&=& {\cal L}_{\mathrm{kin}}[\varphi,\Psi_{v}^{i}] + {\cal L}_{\mathrm{int}}[\varphi,\Psi_{v}^{i}],
\label{eq:Lagrangian_Sigmac_LO_4a}
\end{eqnarray}
with the kinetic term
\begin{eqnarray}
{\cal L}_{\mathrm{kin}}[\varphi,\Psi_{v}^{i}]
&=& \varphi^{\dag}i\frac{\partial}{\partial t}\varphi
  + \varphi^{\dag} \frac{(i\vec{\nabla})^{2}}{2m}\varphi
  + \bar{\Psi}_{v}^{i}i\frac{\partial}{\partial t}\Psi_{v}^{i}
+ {\cal O}(1/M),
\label{eq:Lagrangian_Sigmac_LO_4a_kin}
\end{eqnarray}
and interaction term 
\begin{eqnarray}
{\cal L}_{\mathrm{int}}[\varphi,\Psi_{v}^{i}]
&=&
   C_{1}
   \varphi^{\dag} (\vec{1}_{2} \otimes \vec{1}_{2}) \varphi
            \, \bar{\Psi}_{v}^{i} (\delta^{ij} \otimes \vec{1}_{2} \otimes \vec{1}_{3}) \Psi_{v}^{j}
+ C_{2}
   \varphi^{\dag} (\sigma^{\ell} \otimes \vec{1}_{2}) \varphi
            \, \bar{\Psi}_{v}^{i} (i\varepsilon^{ij\ell} \otimes \vec{1}_{2} \otimes \vec{1}_{3}) \Psi_{v}^{j}
            \nonumber \\ && 
+ C_{3}
   \varphi^{\dag} (\vec{1}_{2} \otimes \tau^{d}) \varphi
            \, \bar{\Psi}_{v}^{i} (\delta^{ij} \otimes \vec{1}_{2} \otimes t^{d}) \Psi_{v}^{j}
+ C_{4}
   \varphi^{\dag} (\sigma^{\ell} \otimes \tau^{d}) \varphi
            \, \bar{\Psi}_{v}^{i} (i\varepsilon^{ij\ell} \otimes \vec{1}_{2} \otimes t^{d}) \Psi_{v}^{j}
+ {\cal O}(1/M),
\label{eq:Lagrangian_Sigmac_LO_4a_int}
\end{eqnarray}
with the coupling constants $C_{A}$ ($A=1,2,3,4$).
We notice that the index $\mu$ in $\Psi_{v}^{\mu}$ is restricted to $i=1,2,3$ in the rest frame.
The above Lagrangian is invariant under the spin symmetry and the isospin symmetry, $\mathrm{SU}(2)_{\mathrm{spin}} \times \mathrm{SU}(2)_{\mathrm{isospin}}$.
In the operator $A \otimes B$ acting on the nucleon ($\varphi$), $A$ and $B$ are the operators for the spin and the isospin of a nucleon.
Similarly, in the operator $A \otimes B \otimes C$ acting on the $\Sigma_{c}$ ($\Sigma_{c}^{\ast}$) baryon ($\Psi_{v}^{i}$), $A$ and $B$ are the operators for the spin of the light component ($qq$) and the spin of the heavy quark ($Q$), respectively, and $C$ is the operator for the isospin of the light component ($qq$).
$\vec{1}_{2}$ is the two-by-two identity matrix for spin or isospin, and $\vec{1}_{3}$ is the three-by-three identity matrix for isospin.
We also use the notations $\sigma^{\ell}$ ($\ell=1,2,3$) and $\tau^{d}$ ($d=1,2,3$) for the Pauli matrices acting on the spin of a nucleon and the isospin of a nucleon or a $\Sigma_{c}$ ($\Sigma_{c}^{\ast}$) baryon, respectively.
We define $\varepsilon^{ij\ell}$ ($\varepsilon^{123}=1$; $i,j,\ell=1,2,3$) as the anti-symmetric tensor for the spin of a $\Sigma_{c}$ baryon or a $\Sigma_{c}^{\ast}$ baryon, and $t^{d}$ ($d=1,2,3$) as the operator for the isospin of a $\Sigma_{c}$ ($\Sigma_{c}^{\ast}$) baryon, whose explicit forms are given by
\begin{eqnarray}
   t^{1}
=
\left(
\begin{array}{ccc}
 0 & 0 & 0 \\
 0 & 0 & -i \\ 
 0 & i & 0
\end{array}
\right),
\hspace{1em}
   t^{2}
=
\left(
\begin{array}{ccc}
 0 & 0 & i \\
 0 & 0 & 0 \\ 
 -i & 0 & 0
\end{array}
\right),
\hspace{1em}
   t^{3}
=
\left(
\begin{array}{ccc}
 0 & -i & 0 \\
 i & 0 & 0 \\ 
 0 & 0 & 0
\end{array}
\right).
\label{eq:t_def}
\end{eqnarray}
They satisfy the following relations: 
\begin{eqnarray}
   \sum_{\rho=1,2,3} (t^{d})_{\mu\rho} (t^{e})_{\rho\nu} = \delta^{d}_{\mu}\delta^{e}_{\nu} - \delta^{d}_{\nu}\delta^{e}_{\mu},
\end{eqnarray}
and this will be used in later calculations.\footnote{Notice the relation $(t^{a})_{\mu\nu}=-i\varepsilon^{a\mu\nu}$.}
With the basis in the isospin operator $t^{a}$, the isospin components in $\Psi_{v1/2}$ and $\Psi_{v3/2}^{\mu}$ are expressed as
\begin{eqnarray}
   \Psi_{v1/2}
=
\left(
\begin{array}{c}
 \frac{-i}{\sqrt{2}} \bigl( \Sigma_{c}^{++} + \Sigma_{c}^{0} \bigr) \\
 \frac{1}{\sqrt{2}} \bigl( \Sigma_{c}^{++} - \Sigma_{c}^{0} \bigr) \\
 -i \Sigma_{c}^{+}
\end{array}
\right),
\hspace{1em} 
   \Psi_{v3/2}^{\mu}
=
\left(
\begin{array}{c}
 \frac{-i}{\sqrt{2}} \bigl( \Sigma_{c}^{\ast++} + \Sigma_{c}^{\ast0} \bigr) \\
 \frac{1}{\sqrt{2}} \bigl( \Sigma_{c}^{\ast++} - \Sigma_{c}^{\ast0} \bigr) \\
 -i \Sigma_{c}^{\ast+}
\end{array}
\right).
\end{eqnarray}
We notice that this representation is not diagonal in the charge basis.
The transformation to the diagonal form by the unitary transformation is shown in the Appendix \ref{sec:spin_charge}.
It is apparent that the Lagrangian \eqref{eq:Lagrangian_Sigmac_LO_4a} has the spin symmetry and the isospin symmetry, $\mathrm{SU}(2)_{\mathrm{spin}} \times \mathrm{SU}(2)_{\mathrm{isospin}}$ for both a nucleon and for a $\Sigma_{c}$ ($\Sigma_{c}^{\ast}$) baryon.
Although the numerical values of the coupling constants $C_{A}$ ($A=1,2,3,4$) have not been known,
 the discussion about the Kondo effect can proceed without the information about the specific value of $C_{A}$ as it will be presented later.

For later convenience, we rewrite the interaction term of Eq.~\eqref{eq:Lagrangian_Sigmac_LO_4a} in a compact form as
\begin{eqnarray}
 {\cal L}_{\mathrm{int}}[\varphi,\Psi_{v}^{i}]
=
   C_{1}
   \varphi^{\dag} \Gamma \varphi \, \bar{\Psi}_{v}^{i} \tilde{\Gamma}_{ij} \Psi_{v}^{j}
+ C_{2}
   \varphi^{\dag}\Gamma^{\ell}\varphi \, \bar{\Psi}_{v}^{i} \tilde{\Gamma}^{\ell}_{ij} \Psi_{v}^{j}
+ C_{3}
   \varphi^{\dag}\Gamma^{d}\varphi \, \bar{\Psi}_{v}^{i} \tilde{\Gamma}^{d}_{ij} \Psi_{v}^{j}
+ C_{4}
   \varphi^{\dag}\Gamma^{\ell d}\varphi \, \bar{\Psi}_{v}^{i} \tilde{\Gamma}^{\ell d}_{ij} \Psi_{v}^{j}
+ {\cal O}(1/M),
 \label{eq:Lagrangian_Sigmac_LO_4}
\end{eqnarray}
where we introduce the following operators:
\begin{eqnarray}
 \Gamma \equiv \vec{1}_{2} \otimes \vec{1}_{2}, \hspace{1em}
 \Gamma^{\ell} \equiv \sigma^{\ell} \otimes \vec{1}_{2}, \hspace{1em}
 \Gamma^{d} \equiv \vec{1}_{2} \otimes \tau^{d}, \hspace{1em}
 \Gamma^{\ell d} \equiv \sigma^{\ell} \otimes \tau^{d},
\label{eq:Gamma_def_4}
\end{eqnarray}
for a nucleon ($\varphi$) and 
\begin{eqnarray}
 \tilde{\Gamma}_{ij} \equiv \delta^{ij} \otimes \vec{1}_{2} \otimes \vec{1}_{3}, \hspace{1em}
 \tilde{\Gamma}^{\ell}_{ij} \equiv i\varepsilon^{ij\ell} \otimes \vec{1}_{2} \otimes \vec{1}_{3}, \hspace{1em}
 \tilde{\Gamma}^{d}_{ij} \equiv \delta^{ij} \otimes \vec{1}_{2} \otimes t^{d}, \hspace{1em}
 \tilde{\Gamma}^{\ell d}_{ij} \equiv i\varepsilon^{ij\ell} \otimes \vec{1}_{2} \otimes t^{d},
\label{eq:tildeGamma_def_4}
\end{eqnarray}
for a $\Sigma_{c}$ ($\Sigma_{c}^{\ast}$) baryon ($\Psi_{v}^{i}$).
The sum is taken over the repeated indices.

Several comments are in order.
First, the heavy-quark spin does not flip by the interaction with a nucleon in the HQS symmetry, and hence we have only the identity operator ($\vec{1}_{2}$) for the heavy quark.
This is because the spin for the heavy quark ($c$ quark) in the $\Sigma_{c}$ ($\Sigma_{c}^{\ast}$) is
 independent of the spin for the light diquark ($qq$).
Thus, to be precise, the total symmetry should be given by $\mathrm{SU}(2)_{\mathrm{light\,spin}}  \times \mathrm{SU}(2)_{\mathrm{heavy\,spin}}  \times \mathrm{SU}(2)_{\mathrm{isospin}}$ including 
$\mathrm{SU}(2)_{\mathrm{heavy\,spin}}$ for the spin symmetry of the heavy quark. 

Second, we remark that the propagator of the nucleon with an energy $p_{0}$ and a three-dimensional momentum $\vec{p}$ in nuclear matter with the chemical potential $\mu$ is given by
\begin{eqnarray}
 \frac{i}{p_{0}-(E_{\vec{p}}-\mu)+i\varepsilon'}
=
 \frac{i\theta(E_{\vec{p}}-\mu)}{p_{0}-(E_{\vec{p}}-\mu)+i\varepsilon}
+ \frac{i\theta(\mu-E_{\vec{p}})}{p_{0}-(E_{\vec{p}}-\mu)-i\varepsilon},
\label{eq:propagator_nucleon_2}
\end{eqnarray}
with $\varepsilon>0$ an infinitely small number.
$E_{\vec{p}}=\vec{p}^{2}/(2m)$ is the energy of the nucleon with a mass $m$, and $\mu$ is the chemical potential for the nucleon.
Notice the difference in the pole positions between the particle component ($E_{\vec{p}}>\mu$) and the hole component ($E_{\vec{p}}<\mu$).
The propagators of the $\Sigma_{c}$ and $\Sigma_{c}^{\ast}$ baryons with an energy $p_{0}$ are given by
\begin{eqnarray}
   \frac{i\delta_{\alpha\beta}}{p_{0}+i\varepsilon}, \hspace{1em}
   \frac{i\delta_{\alpha\beta}\delta^{ij}}{p_{0}+i\varepsilon},
\end{eqnarray}
in rest frame.
Notice that the energy in the denominator ($p_{0}$) describes the residual momentum of the $\Sigma_{c}$ and $\Sigma_{c}^{\ast}$ baryons.

Third, we remark that the $\Sigma_{c}$ baryon and the $\Sigma_{c}^{\ast}$ baryon can decay via $\Sigma_{c} \rightarrow \Lambda_{c}\pi$ and $\Sigma_{c}^{\ast} \rightarrow \Lambda_{c}\pi$, whose decay widths are around 2 MeV and 15 MeV, respectively~\cite{Tanabashi:2018oca}.
In the present study, we consider that the $\Sigma_{c}$ and $\Sigma_{c}^{\ast}$ baryons are in the quasistable states whose lifetimes are long enough.
We also neglect the coupling between the $\Sigma_{c}N$ ($\Sigma_{c}^{\ast}N$) and the $\Lambda_{c}N$ state.
Those subjects are left for future work.

\section{Renormalization group equation}
\label{sec:RG_eq_Sigmac}

In the Kondo effect, the coupling constants in the medium are enhanced logarithmically
 in the low-energy region, and the system becomes a strongly-coupled one.
In this situation, the coupling constants are not the constant values literally, but they should be regarded as the {\it effective coupling constants} whose property is dependent on the relevant energy scale in the medium.
We therefore study how the coupling constant $C_{A}$ ($A=1,2,3,4$) in Eq.~\eqref{eq:Lagrangian_Sigmac_LO_4} is changed into the effective coupling constants in terms of the Kondo effect.
We use the renormalization group (RG) equation.

\subsection{Brief review of RG equation in the Kondo effect}
\label{sec:Kondo_one}

\begin{figure}[tb]
\begin{center}
\vspace{0cm}
\includegraphics[scale=0.35]{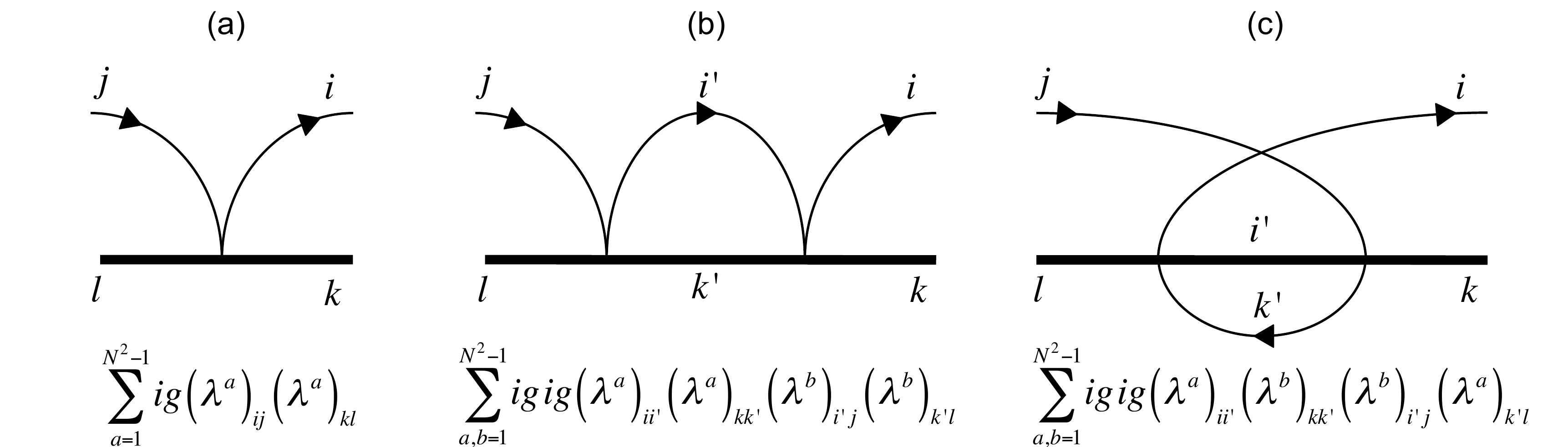}
\vspace{0em}
\caption{The diagrams of the effective interactions between an itinerant fermion (thin lines) and an impurity particle (thick lines) in the $\mathrm{SU}(N)$ interaction. (a) Leading-order term.  (b), (c) Next-to-leading terms (the loop-integral terms). Panels (b) and (c) indicate the particle state and the hole state, respectively, in the intermediate states.}
\label{fig:Fig_190124_simple}
\end{center}
\end{figure}

We begin by briefly reviewing the derivation of the RG equation for the Kondo effect at a simple setting.
As usual, we follow the poor man's scaling proposed by Anderson in Ref.~\cite{Anderson1970} as a standard procedure. 
Treating the impurity particle as an infinitely massive one, 
we consider the Kondo effect with one single non-Abelian interaction between an itinerant fermion (whose mass is $m$) in the Fermi gas and the heavy impurity.
We suppose that they belong to the fundamental representation of the $\mathrm{SU}(N)$ symmetry, and the interaction between the itinerant fermion and the heavy impurity is provided by the non-Abelian interaction
$\displaystyle {\cal L}_{\mathrm{int}}=g(\lambda^{a})_{ij}(\lambda^{a})_{k\ell}$
with the coupling constant $g$ and the Gell-Mann matrices $\lambda^{a}$ ($a=1,2,\dots,N^{2}-1$) in the $\mathrm{SU}(N)$ symmetry.
For example, the case of $N=2$ is the spin-exchange interaction, in which an attraction between the itinerant fermion and the heavy impurity is provided in the spin-antiparallel channel for $g<0$ and in the spin-parallel channel for $g>0$.

Let us consider the scattering amplitude for the itinerant fermion and the impurity.
In the perturbation, the scattering amplitude is given at the tree level at the leading order, and it is affected with particles (nucleons) and holes (nucleon-holes) in the intermediate states in the loops at the next-to-leading order.
The latter contribution supplies the logarithmic divergence at the infrared energy scale near the Fermi surface, and leads to the breakdown of the perturbative treatment.
In order to solve this problem, we consider the RG equation by resumming the logarithmic contributions for the coupling constant.
Following Ref.~\cite{Anderson1970}, we divide the energy of the virtually excited particles and holes into the small pieces.
We then introduce the energy scale $\Lambda$, which is measured from the Fermi energy, and examine how the effective coupling constants are affected by the small modification of $\Lambda$.
We estimate the coupling constants on the lower-energy scale $\Lambda-\drm \Lambda$ by including the loop effect of the particle-hole creations with the energy-shell between $\Lambda-\drm \Lambda$ and $\Lambda$.
The initial value of the coupling constant starting in the RG equation is given by the bare coupling constant in vacuum, $g$.

We denote the interaction coupling at the energy scale $\Lambda$ by $g(\Lambda)$.
Considering the diagrams at tree-level and one-loop in Fig.~\ref{fig:Fig_190124_simple}, we obtain the renormalization group equation for $g(\Lambda)$,
\begin{eqnarray}
&&
   \sum_{a=1}^{N^{2}-1}
   ig(\Lambda-\drm \Lambda)
   \bigl( \lambda^{a} \bigr)_{ij}
   \bigl( \lambda^{a} \bigr)_{kl}
\nonumber \\ 
&=&
   \sum_{a=1}^{N^{2}-1}
   ig(\Lambda)
   \bigl( \lambda^{a} \bigr)_{ij}
   \bigl( \lambda^{a} \bigr)^{kl}
   \nonumber \\ && 
+ \sum_{a,b=1}^{N^{2}-1}
  ig(\Lambda) ig(\Lambda)
   \sum_{i',k'=1}^{N}
   \bigl( \lambda^{a} \bigr)_{ii'} \bigl( \lambda^{a} \bigr)_{kk'} \bigl( \lambda^{b} \bigr)_{i'j} \bigl( \lambda^{b} \bigr)_{k'l}
   \int_{\mathrm{shell}} \frac{\mathrm{d}p_{0}}{2\pi} \frac{\mathrm{d}^{3}\vec{p}}{(2\pi)^{3}}
   \frac{i}{p_{0}-(E_{\vec{p}}-\mu)+i\varepsilon'}
   \frac{i}{-p_{0}+i\varepsilon}
   \nonumber \\ && 
+ \sum_{a,b=1}^{N^{2}-1}
   ig(\Lambda) ig(\Lambda)
   \sum_{i',k'=1}^{N}
   \bigl( \lambda^{a} \bigr)_{ii'} \bigl( \lambda^{b} \bigr)_{kk'} \bigl( \lambda^{b} \bigr)_{i'j} \bigl( \lambda^{a} \bigr)_{k'l}
   \int_{\mathrm{shell}} \frac{\mathrm{d}p_{0}}{2\pi} \frac{\mathrm{d}^{3}\vec{p}}{(2\pi)^{3}}
   \frac{i}{p_{0}-(E_{\vec{p}}-\mu)+i\varepsilon'}
   \frac{i}{p_{0}+i\varepsilon},
\label{eq:RG_simple}
\end{eqnarray}
where $E=\vec{p}^{2}/(2m)$, $\varepsilon$ is a small positive number, and $\varepsilon=\varepsilon$ for $p_{0}>E_{\vec{p}}-\mu$ (particle) and $\varepsilon=-\varepsilon$ for $p_{0}<E_{\vec{p}}-\mu$ (hole).
$\mu$ is the chemical potential for the itinerant fermions.
Considering that the integral region for the momentum is limited to the energy-shell, $|E_{\vec{p}}-\mu| \in [\Lambda-\drm \Lambda,\Lambda]$,
we adopt the following approximations near the Fermi surface:
\begin{eqnarray}
   \int_{\mathrm{shell}} \frac{\mathrm{d}p_{0}}{2\pi} \frac{\mathrm{d}^{3}\vec{p}}{(2\pi)^{3}}
   \frac{i}{p_{0}-(E_{\vec{p}}-\mu)+i\varepsilon'}
   \frac{i}{-p_{0}+i\varepsilon}
&\simeq&
   -i
   \rho_{0} \frac{\drm \Lambda}{\Lambda},
\label{eq:momentum_integral_1}
\end{eqnarray}
and
\begin{eqnarray}
   \int_{\mathrm{shell}} \frac{\mathrm{d}p_{0}}{2\pi} \frac{\mathrm{d}^{3}\vec{p}}{(2\pi)^{3}}
   \frac{i}{p_{0}-(E_{\vec{p}}-\mu)+i\varepsilon'}
   \frac{i}{p_{0}+i\varepsilon}
&\simeq&
   i      
   \rho_{0} \frac{\drm \Lambda}{\Lambda},
\label{eq:momentum_integral_2}
\end{eqnarray}
where we leave only the leading terms for a small $\drm \Lambda/\Lambda \ll 1$.
We introduce $\rho_{0} \equiv m^{3/2}\sqrt{2\mu}/(2\pi^{2})$ for the state-number-density at the Fermi surface.
We also use the relationships for the Gell-Mann matrices,
\begin{eqnarray}
   \sum_{a,b=1}^{N^{2}-1} \sum_{i',k'=1}^{N}
   \bigl( \lambda^{a} \bigr)_{ii'} \bigl( \lambda^{a} \bigr)_{kk'} \bigl( \lambda^{b} \bigr)_{i'j} \bigl( \lambda^{b} \bigr)_{k'l}
&=& 
   4 \biggl(1-\frac{1}{N^{2}}\biggr) \delta_{ij}\delta_{k\ell}
+ \biggl(-\frac{4}{N}\biggr) \sum_{a=1}^{N^{2}-1} \bigl( \lambda^{a}_{ij}) \bigl( \lambda^{a} \bigr)_{kl},
 \nonumber \\
   \sum_{a,b=1}^{N^{2}-1} \sum_{i',k'=1}^{N}
   \bigl( \lambda^{a} \bigr)_{ii'} \bigl (\lambda^{b} \bigr)_{kk'} \bigl( \lambda^{b} \bigr)_{i'j} \bigl( \lambda^{a} \bigr)_{k'l}
&=&
   4 \biggl(1-\frac{1}{N^{2}}\biggr) \delta_{ij}\delta_{k\ell}
+ \biggl(2N-\frac{4}{N}\biggr) \sum_{a=1}^{N^{2}-1} \bigl( \lambda^{a}_{ij} \bigr) \bigl( \lambda^{a} \bigr)_{kl}.
\label{eq:lambda4_Fierz}
\end{eqnarray}
Thus, we simplify Eq.~\eqref{eq:RG_simple} to the simplified form
\begin{eqnarray}
    ig(\Lambda-\mathrm{d}\Lambda)
 =
    ig(\Lambda)
 + ig(\Lambda) ig(\Lambda) \biggl( -\frac{4}{N} \biggr) \biggl( -i\rho_{0}\frac{\mathrm{d}\Lambda}{\Lambda} \biggr)
 + ig(\Lambda) ig(\Lambda) \biggl( 2N-\frac{4}{N} \biggr) \biggl( i\rho_{0}\frac{\mathrm{d}\Lambda}{\Lambda} \biggr).
\end{eqnarray}
Taking the limit $\mathrm{d}\Lambda \rightarrow 0$ for a small strip of the momentum shell, we finally obtain the RG equation
\begin{eqnarray}
   \frac{\drm}{\drm \lambda} g(\lambda)
= -2\rho_{0}N g(\lambda)^{2},
   \label{eq:RG_simple_equation}
\end{eqnarray}
with the energy scale $\lambda=-\ln(\Lambda/\Lambda_{0})$.
Here $\Lambda$ is the energy scale moving from the high-energy to the low-energy region, and $\Lambda_{0}$ is the ultraviolet-energy scale as the initial point.
We emphasize that the the minus sign in the right-hand side in Eq.~\eqref{eq:RG_simple_equation} stems from the coefficients in the identities in Eq.~\eqref{eq:lambda4_Fierz} as the nontrivial factors by the non-Abelian properties of the $\lambda^{a}$ matrices.
The solution of the RG equation \eqref{eq:RG_simple_equation} is found to be
\begin{eqnarray}
   g(\lambda) = \frac{g}{1+2\rho_{0}Ng\lambda},
\end{eqnarray}
with $g$ being the coupling constant in vacuum or in the interaction Lagrangian.
Given that the energy scale runs from $\lambda=0$ (the high-energy scale) to $\lambda\rightarrow\infty$ (the low-energy scale),
we find that the negative coupling constant ($g<0$) leads to divergence of the coupling constant $g(\lambda)$ at $\lambda_{\mathrm{K}}=-1/(2\rho_{0}Ng)$ or $\Lambda_{\mathrm{K}}=e^{1/(2\rho_{0}Ng)}$ and that the positive coupling constant ($g>0$) leads to the vanishing coupling constant ($g(\lambda)\rightarrow0$).
The relevant fixed point in the former case produces the Kondo effect, while the irrelevant fixed point in the latter does not.
Thus, the coupling strength in the spin-antiparallel channel is enhanced, while that in the spin-parallel channel is suppressed.
Therefore, the coupling constant becomes enhanced at the low-energy scale by virtue of the non-Abelian property of the interaction. 

\subsection{RG equation for $\Sigma_{c}$ and $\Sigma_{c}^{\ast}$ baryon}

\begin{figure}[tb]
\begin{center}
\vspace{0cm}
\includegraphics[scale=0.35]{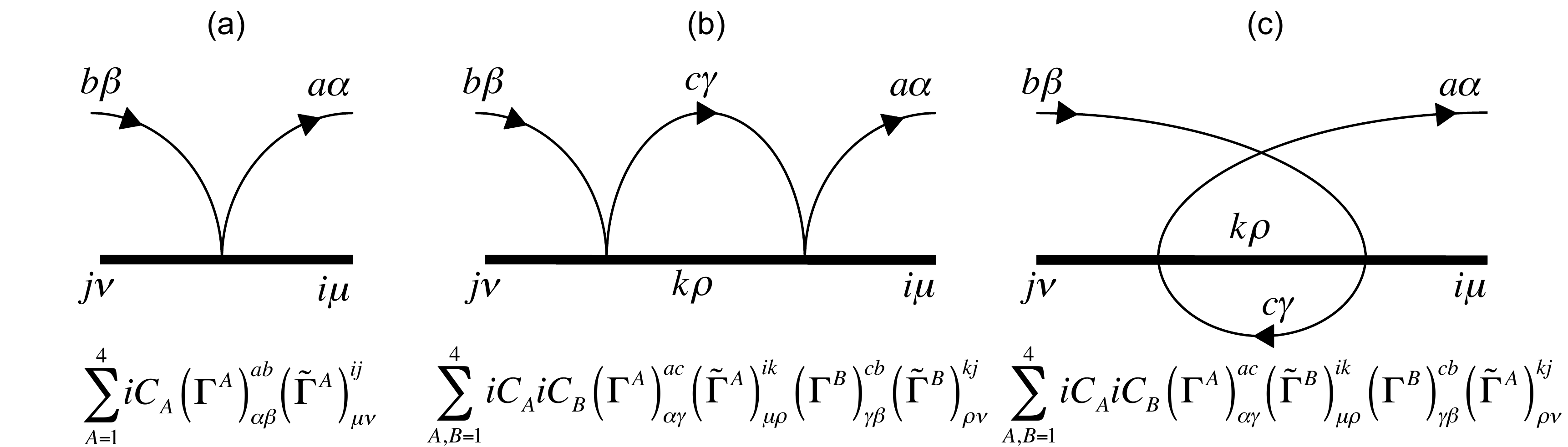}
\vspace{0em}
\caption{The diagrams of the effective interactions between a nucleon and a $\Sigma_{c}$ ($\Sigma_{c}^{\ast}$) baryon in nuclear matter. The thin lines are for the nucleon and the thick lines are for the $\Sigma_{c}$ ($\Sigma_{c}^{\ast}$) baryon. (a) Leading-order term. (b), (c) Next-to-leading terms (the loop-integral terms). Panels (b) and (c) indicate the particle state and the hole state, respectively, in the intermediate states.}
\label{fig:Fig_190124}
\end{center}
\end{figure}

\begin{figure}[tb]
\begin{center}
\vspace{0cm}
    \begin{tabular}{c}
      \begin{minipage}[t]{0.33\hsize}
        \begin{center}
          \raisebox{0em}{\includegraphics[scale=0.33]{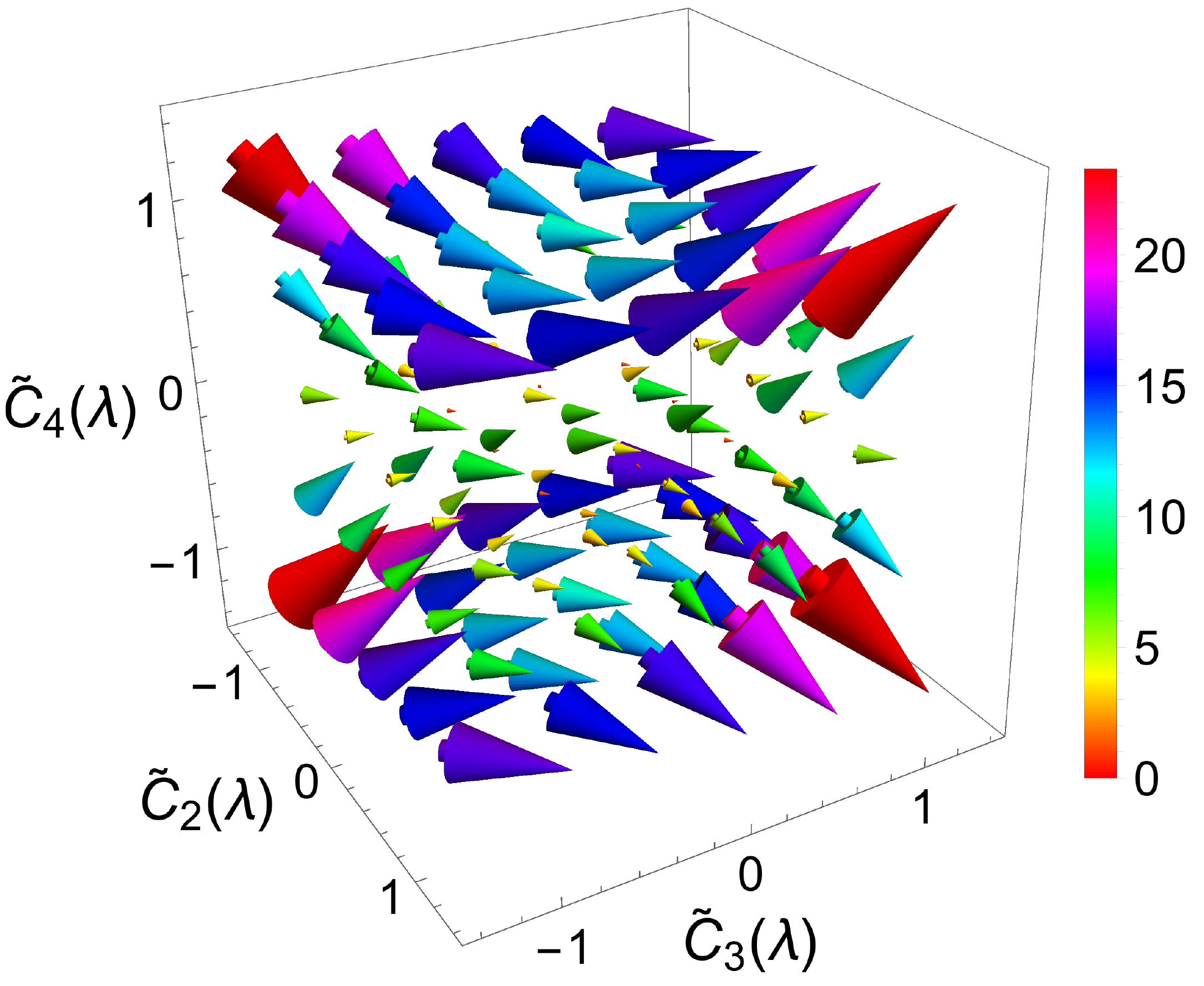}}
        \end{center}
      \end{minipage}
      \hspace{6em}
      \begin{minipage}[t]{0.33\hsize}
      \vspace{0.3em}
        \begin{center}
          \includegraphics[scale=0.24]{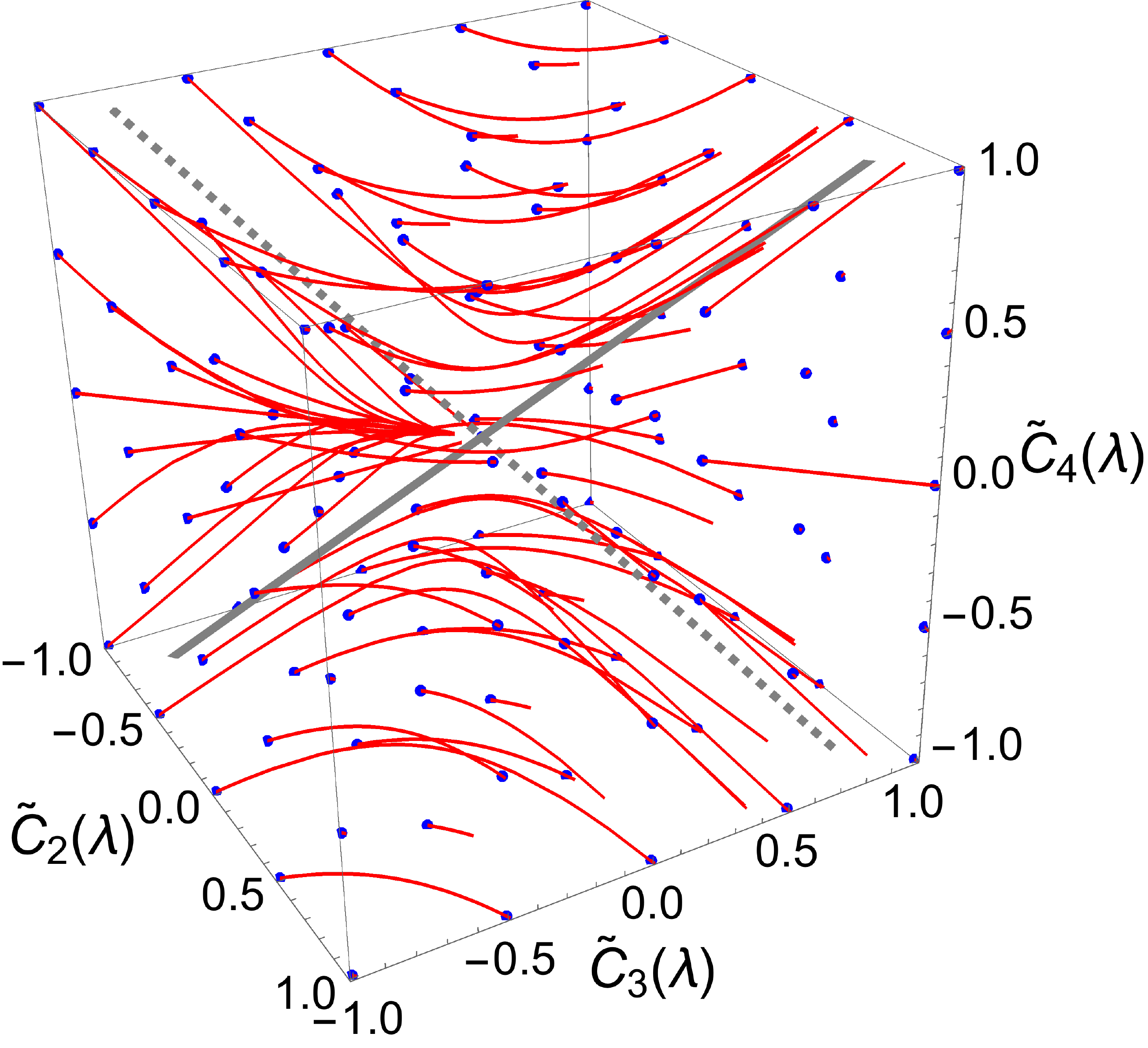}
        \end{center}
      \end{minipage}
    \end{tabular}
\caption{Left: the plot of the vector $\bigl( \tilde{C}_{2}(\lambda)^{2} + 2\tilde{C}_{4}(\lambda)^{2},\tilde{C}_{3}(\lambda)^{2} + 2\tilde{C}_{4}(\lambda)^{2},2 \bigl( \tilde{C}_{2}(\lambda) + \tilde{C}_{3}(\lambda) \bigr) \tilde{C}_{4}(\lambda) \bigr)$, i.e., the right-hand side of Eq.~\eqref{eq:RG_Kondo_Sigma_1} in the three-dimensional parameter space $\bigl( \tilde{C}_{2}(\lambda),\tilde{C}_{3}(\lambda),\tilde{C}_{4}(\lambda) \bigr)$. Right: the stream lines of $\bigl( \tilde{C}_{2}(\lambda),\tilde{C}_{3}(\lambda),\tilde{C}_{4}(\lambda) \bigr)$ as the solutions of Eq.~\eqref{eq:RG_Kondo_Sigma_1}. The initial conditions $(\tilde{C}_{2},\tilde{C}_{3},\tilde{C}_{4})$ are expressed by dots. The solid and dashed lines with gray indicate the manifold in the $\mathrm{SU}(4)$ limit (cf.~Sec.~\ref{sec:SU4_limit}).}
\label{fig:Fig3}
\end{center}
\end{figure}

Now we apply the RG equation for the coupling constants for the nucleon and the $\Sigma_{c}$ and $\Sigma_{c}^{\ast}$ baryon in Eq.~\eqref{eq:Lagrangian_Sigmac_LO_4}, where the relevant energy should be $\Lambda_{0}$  as an initial scale.
Repeating the similar argument in the previous subsection, at the one-loop order, we find that the RG equation reads
\begin{eqnarray}
&&
   \sum_{A=1}^{4}
   iC_{A}(\Lambda-\drm \Lambda)
   \bigl( \Gamma^{A} \bigr)^{ab}_{\alpha\beta}
   \bigl( \tilde{\Gamma}^{A} \bigr)^{ij}_{\mu\nu}
\nonumber \\ 
&=&
   \sum_{A=1}^{4}
   iC_{A}(\Lambda)
   \bigl( \Gamma^{A} \bigr)^{ab}_{\alpha\beta}
   \bigl( \tilde{\Gamma}^{A} \bigr)^{ij}_{\mu\nu}
   \nonumber \\ && 
+ \sum_{A,B=1}^{4}
  iC_{A}(\Lambda) iC_{B}(\Lambda)
   \bigl( \Gamma^{A} \bigr)^{ac}_{\alpha\gamma}
   \bigl( \tilde{\Gamma}^{A} \bigr)^{ik}_{\mu\rho}
   \bigl( \Gamma^{B} \bigr)^{cb}_{\gamma\beta}
   \bigl( \tilde{\Gamma}^{B} \bigr)^{kj}_{\rho\nu}
   \int_{\mathrm{shell}} \frac{\mathrm{d}p_{0}}{2\pi} \frac{\mathrm{d}^{3}\vec{p}}{(2\pi)^{3}}
   \frac{i}{p_{0}-(E_{\vec{p}}-\mu)+i\varepsilon'}
   \frac{i}{-p_{0}+i\varepsilon}
   \nonumber \\ && 
+ \sum_{A,B=1}^{4}
   iC_{A}(\Lambda) iC_{B}(\Lambda)
   \bigl( \Gamma^{A} \bigr)^{ac}_{\alpha\gamma}
   \bigl( \tilde{\Gamma}^{B} \bigr)^{ik}_{\mu\rho}
   \bigl( \Gamma^{B} \bigr)^{cb}_{\gamma\beta}
   \bigl( \tilde{\Gamma}^{A} \bigr)^{kj}_{\rho\nu}
   \int_{\mathrm{shell}} \frac{\mathrm{d}p_{0}}{2\pi} \frac{\mathrm{d}^{3}\vec{p}}{(2\pi)^{3}}
   \frac{i}{p_{0}-(E_{\vec{p}}-\mu)+i\varepsilon'}
   \frac{i}{p_{0}+i\varepsilon},
\label{eq:RG0}
\end{eqnarray}
where
the term on the left-hand side denotes the effective coupling constants on the energy scale $\Lambda-\drm \Lambda$, and,
on the right-hand side, the first term denotes the effective coupling constant on the energy scale $\Lambda$, and the second (third) term denotes the loop integrals with particle (hole) creation in the energy-shell between $\Lambda-\drm \Lambda$ and $\Lambda$ (Fig.~\ref{fig:Fig_190124}).
In the above equation, the indices in the operator $\Gamma^{A}$ and $\tilde{\Gamma}^{A}$ ($A=1,2,3,4$) are shown as
\begin{eqnarray}
   \bigl( \Gamma^{1} \bigr)^{ab}_{\alpha\beta}
= \delta^{ab}\delta_{\alpha\beta}, \hspace{0.5em}
\bigl( \Gamma^{2} \bigr)^{ab}_{\alpha\beta}
= \bigl(\sigma^{\ell}\bigr)^{ab}\delta_{\alpha\beta}, \hspace{0.5em}
\bigl( \Gamma^{3} \bigr)^{ab}_{\alpha\beta}
= \delta^{ab}(\tau^{d})_{\alpha\beta}, \hspace{0.5em}
\bigl( \Gamma^{4} \bigr)^{ab}_{\alpha\beta}
= \bigl(\sigma^{\ell}\bigr)^{ab}\bigl(\tau^{d}\bigr)_{\alpha\beta},
\end{eqnarray}
for the nucleon part,
and
\begin{eqnarray}
   \bigl( \tilde{\Gamma}^{1} \bigr)^{ij}_{\mu\nu}
= \delta^{ij}\delta_{\mu\nu}, \hspace{0.5em}
   \bigl( \tilde{\Gamma}^{2} \bigr)^{ij}_{\mu\nu}
= i\varepsilon^{ij\ell}\delta_{\mu\nu}, \hspace{0.5em}
   \bigl( \tilde{\Gamma}^{3} \bigr)^{ij}_{\mu\nu}
= \delta^{ij}(t^{d})_{\mu\nu}, \hspace{0.5em}
   \bigl( \tilde{\Gamma}^{4} \bigr)^{ij}_{\mu\nu}
= i\varepsilon^{ij\ell}\bigl(t^{d}\bigr)_{\mu\nu},
\end{eqnarray}
for the $\Sigma_{c}$ ($\Sigma_{c}^{\ast}$) baryon part.
Here $a,b=1,2$ and $\alpha,\beta=1,2$ are for the spin and the isospin of a nucleon, respectively,
and $i,j=1,2,3$ and $\mu,\nu=1,2,3$ are for the spin and for the isospin of a diquark component ($qq$) in a $\Sigma_{c}$ ($\Sigma_{c}^{\ast}$) baryon, respectively.
We consider that the sum over the spin direction ($\ell=1,2,3$) and the isospin direction ($d=1,2,3$) is included if necessary.
Utilizing the momentum integrals in Eqs.~\eqref{eq:momentum_integral_1} and \eqref{eq:momentum_integral_2},
we rewrite the RG equation \eqref{eq:RG0} as
\begin{eqnarray}
   \frac{\drm}{\drm \lambda} C_{1}(\lambda)
&=& 0,
\nonumber \\
   \frac{\drm}{\drm \lambda} C_{2}(\lambda)
&=&
   \rho_{0}
   \bigl( 4 C_{2}(\lambda)^{2} + 8C_{4}(\lambda)^{2} \bigr),
\nonumber \\
   \frac{\drm}{\drm \lambda} C_{3}(\lambda)
&=&
    \rho_{0}
   \bigl(
            -4C_{3}(\lambda)^{2} - 8C_{4}(\lambda)^{2}
   \bigr),
\nonumber \\
   \frac{\drm}{\drm \lambda} C_{4}(\lambda)
&=&
   \rho_{0}
   \bigl(
         8C_{2}(\lambda) C_{4}(\lambda) - 8C_{3}(\lambda) C_{4}(\lambda)
   \bigr),
\label{eq:RG1}
\end{eqnarray}
for each channel $A=1,2,3,4$.
Here, we introduce the new variable $\displaystyle \lambda \equiv - \ln \bigl(\Lambda/\Lambda_{0}\bigr)$ instead of the energy scale $\Lambda$.
The high-energy scale $\Lambda_{0}$ for which the RG equation starts is set to be equal to the chemical potential of the nuclear matter $\mu$ or the cutoff energy-scale $D$ in the point-like interaction in Eq.~\eqref{eq:Lagrangian_Sigmac_LO_4}.
In the present discussion, however, there is no necessity to specify the value of $\Lambda_{0}$ explicitly.
We notice that the variable $\lambda$ changes from $\lambda=0$ to $\lambda \rightarrow \infty$ in correspondence to the change from the high-energy scale to the low-energy scale.
As seen in Eq.~\eqref{eq:RG1}, $C_{1}(\lambda)$ is not affected by the change of $\lambda$, and hence the spin and isospin-independent channels are not subject to the medium effect.
Thus, we will consider only $C_{2}(\lambda)$, $C_{3}(\lambda)$, and $C_{4}(\lambda)$ in the following discussions.
For convenience, we use the following dimensionless effective coupling constants:
\begin{eqnarray}
   \tilde{C}_{2}(\lambda) \equiv 4\rho_{0}C_{2}(\lambda), \hspace{1em}
   \tilde{C}_{3}(\lambda) \equiv -4\rho_{0}C_{3}(\lambda), \hspace{1em}
   \tilde{C}_{4}(\lambda) \equiv 4\rho_{0}C_{4}(\lambda),
\label{eq:RG_Kondo_Sigma_0}
\end{eqnarray}
instead of $C_{2}(\lambda)$, $C_{3}(\lambda)$, and $C_{4}(\lambda)$,
and rewrite the RG equation
\eqref{eq:RG1} as
\begin{eqnarray}
   \frac{\drm}{\drm \lambda} \tilde{C}_{2}(\lambda) &=& \tilde{C}_{2}(\lambda)^{2} + 2\tilde{C}_{4}(\lambda)^{2},
\nonumber \\
   \frac{\drm}{\drm \lambda} \tilde{C}_{3}(\lambda) &=& \tilde{C}_{3}(\lambda)^{2} + 2\tilde{C}_{4}(\lambda)^{2},
\nonumber \\
   \frac{\drm}{\drm \lambda} \tilde{C}_{4}(\lambda) &=& 2 \bigl( \tilde{C}_{2}(\lambda) + \tilde{C}_{3}(\lambda) \bigr) \tilde{C}_{4}(\lambda).
\label{eq:RG_Kondo_Sigma_1}
\end{eqnarray}
Those are the basic equations used in the following discussions.
Notice that we have added the minus sign for $C_{3}(\lambda)$ in Eq.~\eqref{eq:RG_Kondo_Sigma_0} simply for the appearance of the equations.
The initial conditions are given as $\tilde{C}_{2}(0)=4\rho_{0}C_{2}$, $\tilde{C}_{3}(0)=-4\rho_{0}C_{3}$, and $\tilde{C}_{4}(0)=4\rho_{0}C_{4}$ with $C_{2}$, $C_{3}$, and $C_{4}$ being the coupling constants in the interaction Lagrangian~\eqref{eq:Lagrangian_Sigmac_LO_4}.
In Fig.~\ref{fig:Fig3},
we plot the right-hand side of Eq.~\eqref{eq:RG_Kondo_Sigma_1}, i.e., the vector $\bigl( \tilde{C}_{2}(\lambda)^{2} + 2\tilde{C}_{4}(\lambda)^{2},\tilde{C}_{3}(\lambda)^{2} + 2\tilde{C}_{4}(\lambda)^{2},2 \bigl( \tilde{C}_{2}(\lambda) + \tilde{C}_{3}(\lambda) \bigr) \tilde{C}_{4}(\lambda) \bigr)$ in the three-dimensional parameter space $\bigl( \tilde{C}_{2}(\lambda),\tilde{C}_{3}(\lambda),\tilde{C}_{4}(\lambda) \bigr)$,
and also show the stream lines for $\bigl( \tilde{C}_{2}(\lambda),\tilde{C}_{3}(\lambda),\tilde{C}_{4}(\lambda) \bigr)$ varying with $\lambda$ and the several initial conditions $(\tilde{C}_{2},\tilde{C}_{3},\tilde{C}_{4})$ at $\lambda=0$ as the solutions of Eq.~\eqref{eq:RG_Kondo_Sigma_1}.
The initial conditions are plotted by the dots in the figure.
We notice that, for the increasing $\lambda$, there are some initial conditions giving the stream lines convergent to zero and the other initial conditions giving the stream lines divergent.
In the following subsections, we will investigate the solutions of Eq.~\eqref{eq:RG_Kondo_Sigma_1} in detail.
We will find that the $C_{4}$ term, i.e., the spin and isospin-dependent term in Eq.~\eqref{eq:Lagrangian_Sigmac_LO_4} plays an important role to extend the parameter region of the coupling constants in which the Kondo effect occurs.

\subsection{Analytical solutions in special cases}

Although Eq.~\eqref{eq:RG_Kondo_Sigma_1} may look simple,
it is difficult to obtain the analytic solution due to the nonlinearity of the equation.
Therefore, we have to perform the numerical calculation.
In order to understand roughly the properties of the solutions before the numerical computing,
we seek to obtain analytic solutions by restricting the parameter space to simpler subspaces and focusing on special cases:
(i) $\tilde{C}_{4}(\lambda)=0$,
(ii) $\tilde{C}_{3}(\lambda)=0$ (or $\tilde{C}_{2}(\lambda)=0$),
(iii) $\tilde{C}_{2}(\lambda)=\tilde{C}_{3}(\lambda)$ with $|\tilde{C}_{4}(\lambda)| \ll 1$, and
(iv) $\displaystyle \tilde{C}_{2}(\lambda)=\tilde{C}_{3}(\lambda)=\pm\sqrt{2/3}\,\tilde{C}_{4}(\lambda)$.
We will show that, in the last case, the $\mathrm{SU}(4)$ symmetry is realized as an extension from the $\mathrm{SU}(2)_{\mathrm{spin}} \times \mathrm{SU}(2)_{\mathrm{isospin}}$ symmetry in the Lagrangian.
The simple settings from (i) to (iv) will provide us with fresh insights about the Kondo effect for the $\Sigma_{c}$ ($\Sigma_{c}^{\ast}$) baryon in the nuclear matter.

\subsubsection{Conventional case}
\label{sec:C40}

We consider the case of $\tilde{C}_{4}(\lambda)=0$ ($C_{4}=0$), i.e., neglecting the spin and isospin-dependent term in the interaction.
Then, the RG equation~\eqref{eq:RG_Kondo_Sigma_1} is transformed to
\begin{eqnarray}
   \frac{\drm}{\drm \lambda} \tilde{C}_{2}(\lambda) &=& \tilde{C}_{2}(\lambda)^{2},
\nonumber \\
   \frac{\drm}{\drm \lambda} \tilde{C}_{3}(\lambda) &=& \tilde{C}_{3}(\lambda)^{2},
\nonumber \\
   \frac{\drm}{\drm \lambda} \tilde{C}_{4}(\lambda) &=& 0.
\label{eq:RG_Kondo_Sigma_2}
\end{eqnarray}
We find that $\tilde{C}_{4}(\lambda)$ is constant, while $\tilde{C}_{2}(\lambda)$ and $\tilde{C}_{3}(\lambda)$ change according to the change of the energy scale.
Because $\tilde{C}_{2}(\lambda)$ and $\tilde{C}_{3}(\lambda)$ are decoupled,
 the spin-dependent term and the isospin-dependent term 
  obeys the usual Kondo effect with a single non-Abelian symmetry.
  The Kondo effect of the single non-Abelian symmetry is summarized in Sec.~\ref{sec:Kondo_one}.
The solutions of $\tilde{C}_{2}(\lambda)$ and $\tilde{C}_{3}(\lambda)$ are found to be
\begin{eqnarray}
   \tilde{C}_{2}(\lambda) &=& \frac{\tilde{C}_{2}}{1-\tilde{C}_{2}\lambda}, \nonumber \\ 
   \tilde{C}_{3}(\lambda) &=& \frac{\tilde{C}_{3}}{1-\tilde{C}_{3}\lambda},
\end{eqnarray}
with $\tilde{C}_{2}=4\rho_{0}C_{2}$ and $\tilde{C}_{3}=-4\rho_{0}C_{3}$ as the initial condition at $\lambda=0$.
Thus, the three-dimensional parameter space is essentially reduced to the one-dimensional one.
Let us consider the behavior of the solution $\tilde{C}_{2}(\lambda)$ in detail
in the energy scales from $\lambda=0$ (high energy) to a larger value (low energy).
For the positive value of $C_{2}$ ($C_{2}>0$),
we notice that $\tilde{C}_{2}(\lambda)$ becomes divergent at the end of the energy scale 
$\Lambda=\Lambda_{\mathrm{K}}$ with $\Lambda_{\mathrm{K}}=\Lambda_{0}e^{-1/(4\rho_{0}C_{2})}$.
$\Lambda_{\mathrm{K}}$ is called the Kondo scale (the Landau pole) whose quantity is smaller than $\Lambda_{0}$ due to the exponential factor.
At the Kondo scale, the coupling constant becomes sufficiently large. 
Thus, the system becomes a strongly coupled one and the nonperturbative analysis should be adopted.
For the negative value of $C_{2}$ ($C_{2}<0$), the effective coupling constant becomes zero without divergence in the low-energy limit ($\lambda \rightarrow \infty$), and hence such interaction disappears in the ground state.

A similar analysis is applied to the case of $\tilde{C}_{3}(\lambda)$.
We find that the effective coupling constant becomes divergent at the Kondo scale $\Lambda_{\mathrm{K}}'=\Lambda_{0}e^{1/(4\rho_{0}C_{3})}$ for the negative value of $C_{3}$ ($C_{3}<0$), while it disappears for the positive value of $C_{3}$ ($C_{3}>0$).
Notice that the sign of $C_{3}$ for the Kondo effect is different from that of $C_{2}$ due to the definition in Eq.~\eqref{eq:RG_Kondo_Sigma_0} and that the values of $\Lambda_{\mathrm{K}}$ and $\Lambda_{\mathrm{K}}'$ can be different in general.

So far we have set $\tilde{C}_{4}(\lambda)=0$ ($C_{4}=0$) by neglecting the spin and isospin-dependent term in Eq.~\eqref{eq:Lagrangian_Sigmac_LO_4}, and have seen that $C_{2}<0$ and $C_{3}>0$ lead to the absence of the Kondo effect.
However, this is the case only for $\tilde{C}_{4}(\lambda)=0$ ($C_{4}=0$).
In the following cases, we will demonstrate that the Kondo effect can occur even for $C_{2}<0$ and $C_{3}>0$ when a nonzero value of $\tilde{C}_{4}(\lambda)$ is considered.

\subsubsection{Two-dimensional case I}

By setting $\tilde{C}_{3}(\lambda)=0$  in Eq.~\eqref{eq:RG_Kondo_Sigma_1}, we consider the two-dimensional parameter space spanned by $(\tilde{C}_{2}(\lambda),\tilde{C}_{4}(\lambda))$.
We present the case of $\tilde{C}_{3}(\lambda)=0$ for the demonstration.
A similar conclusion is reached also for $(\tilde{C}_{3}(\lambda),\tilde{C}_{4}(\lambda))$ by setting $\tilde{C}_{2}(\lambda)=0$.
By setting $\tilde{C}_{3}(\lambda)=0$, the RG equation \eqref{eq:RG_Kondo_Sigma_1} is reduced to
\begin{eqnarray}
   \frac{\drm}{\drm \lambda} \tilde{C}_{2}(\lambda) &=& \tilde{C}_{2}(\lambda)^{2} + 2\tilde{C}_{4}(\lambda)^{2},
\nonumber \\
   \frac{\drm}{\drm \lambda} \tilde{C}_{4}(\lambda) &=& 2\tilde{C}_{2}(\lambda) \tilde{C}_{4}(\lambda).
\label{eq:Sigma_c_Kondo_RG_i0}
\end{eqnarray}
To find the solution, we eliminate $\tilde{C}_{2}(\lambda)$ in the above equations, and obtain the equation for $\tilde{C}_{4}(\lambda)$,
\begin{eqnarray}
     \tilde{C}_{4}(\lambda) \frac{\drm^{2}}{\drm \lambda^{2}} \tilde{C}_{4}(\lambda)
   - \frac{3}{2} \biggl( \frac{\drm}{\drm \lambda} \tilde{C}_{4}(\lambda) \biggr)^{2}
  - 4\tilde{C}_{4}(\lambda)^{4}
=0.
\label{eq:Sigma_c_Kondo_RG_i}
\end{eqnarray}
Interestingly, this nonlinear differential equation has a simple analytical solution.
As a result we obtain the solutions
\begin{eqnarray}
   \tilde{C}_{2}(\lambda)
&=&
   \frac
   {\bigl(-\tilde{C}_{2}^{2}+2\tilde{C}_{4}^{2}\bigr)\lambda+\tilde{C}_{2}}
   {1-2\tilde{C}_{2}\lambda+\bigl(\tilde{C}_{2}^{2}-2\tilde{C}_{4}^{2}\bigr)\lambda^{2}},
\nonumber \\ 
   \tilde{C}_{4}(\lambda)
&=&
   \frac
   {\tilde{C}_{4}}
   {1-2\tilde{C}_{2}\lambda+\bigl(\tilde{C}_{2}^{2}-2\tilde{C}_{4}^{2}\bigr)\lambda^{2}},
\label{eq:Sigma_c_Kondo_RG_i_sol}
\end{eqnarray}
with $\tilde{C}_{2}=4\rho_{0}C_{2}$ and $\tilde{C}_{4}=4\rho_{0}C_{4}$ as the initial condition.
The Kondo effect occurs when $\tilde{C}_{2}(\lambda)$ and $\tilde{C}_{4}(\lambda)$ become divergent at a large value of $\lambda$ as the Kondo scale.
To find the Kondo scale,
 we solve
$\bigl(\tilde{C}_{2}^{2}-2\tilde{C}_{4}^{2}\bigr)\lambda^{2}-2\tilde{C}_{2}\lambda+1=0$,
and 
we obtain $\lambda=\lambda_{\pm}$ with
$\displaystyle \lambda_{\pm}=1/\bigl(\tilde{C}_{2} \pm \sqrt{2}\tilde{C}_{4}\bigr)$.
In order for that either $\lambda_{+}>0$ or $\lambda_{-}>0$ is satisfied, 
the values of $\tilde{C}_{2}$ and $\tilde{C}_{4}$ should satisfy $\displaystyle \tilde{C}_{4}>-\tilde{C}_{2}/\sqrt{2}$ or $\displaystyle \tilde{C}_{4}<\tilde{C}_{2}/\sqrt{2}$ in the two-dimensional parameter space $(\tilde{C}_{2},\tilde{C}_{4})$.
The Kondo scale $\Lambda_{\mathrm{K}}$ is obtained as
\begin{eqnarray}
   \Lambda_{\mathrm{K}}
&=&
   \Lambda_{0}
   \exp
   \left(
       - \frac{1}
         {4\rho_{0}\,
         \mathrm{max}
               \bigl(
                     C_{2} + \sqrt{2}C_{4},
                     C_{2} - \sqrt{2}C_{4}
              \bigr)
        }
   \right)
   \hspace{0.5em}\mathrm{for}\hspace{0.5em}
   \tilde{C}_{4}>-\tilde{C}_{2}/\sqrt{2}
   \hspace{0.5em}\mathrm{and}\hspace{0.5em}   
   \tilde{C}_{4}<\tilde{C}_{2}/\sqrt{2},
\nonumber \\
   \Lambda_{\mathrm{K}}
&=&
   \Lambda_{0}
   \exp
   \left(
       - \frac{1}
         {4\rho_{0} \bigl(C_{2} + \sqrt{2}C_{4}\bigr)}
   \right)
   \hspace{0.5em}\mathrm{for}\hspace{0.5em}
   \tilde{C}_{4}>-\tilde{C}_{2}/\sqrt{2}
   \hspace{0.5em}\mathrm{and}\hspace{0.5em}   
   \tilde{C}_{4}<\tilde{C}_{2}/\sqrt{2},
\nonumber \\
   \Lambda_{\mathrm{K}}
&=&
   \Lambda_{0}
   \exp
   \left(
       - \frac{1}
         {4\rho_{0} \bigl(C_{2} - \sqrt{2}C_{4}\bigr)}
   \right)
   \hspace{0.5em}\mathrm{for}\hspace{0.5em}
   \tilde{C}_{4}<-\tilde{C}_{2}/\sqrt{2}
   \hspace{0.5em}\mathrm{and}\hspace{0.5em}   
   \tilde{C}_{4}<\tilde{C}_{2}/\sqrt{2},
\end{eqnarray}
with $\Lambda_{0}$ being the high-energy scale ($\mu$ or $D$) as the initial condition.
The equation forms of the Kondo scale are dependent on the region of $(\tilde{C}_{2},\tilde{C}_{4})$.

\begin{figure}[tb]
\begin{center}
    \begin{tabular}{c}
      \begin{minipage}[t]{0.33\hsize}
        \begin{center}
          \raisebox{0em}{\includegraphics[scale=0.25]{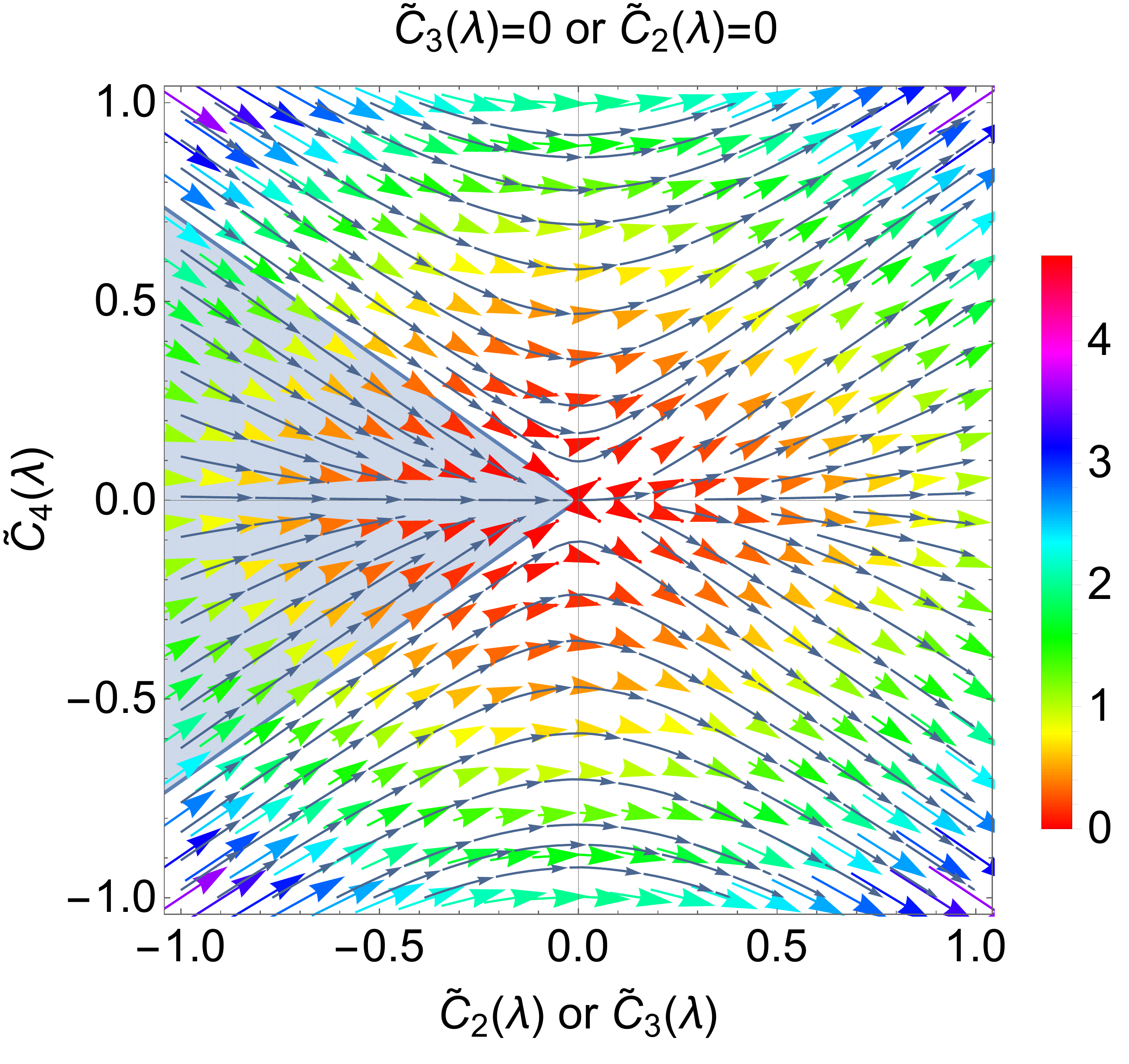}}
        \end{center}
      \end{minipage}
      \hspace{4em}
      \begin{minipage}[t]{0.33\hsize}
      \vspace{-0.6em}
        \begin{center}
          \includegraphics[scale=0.265]{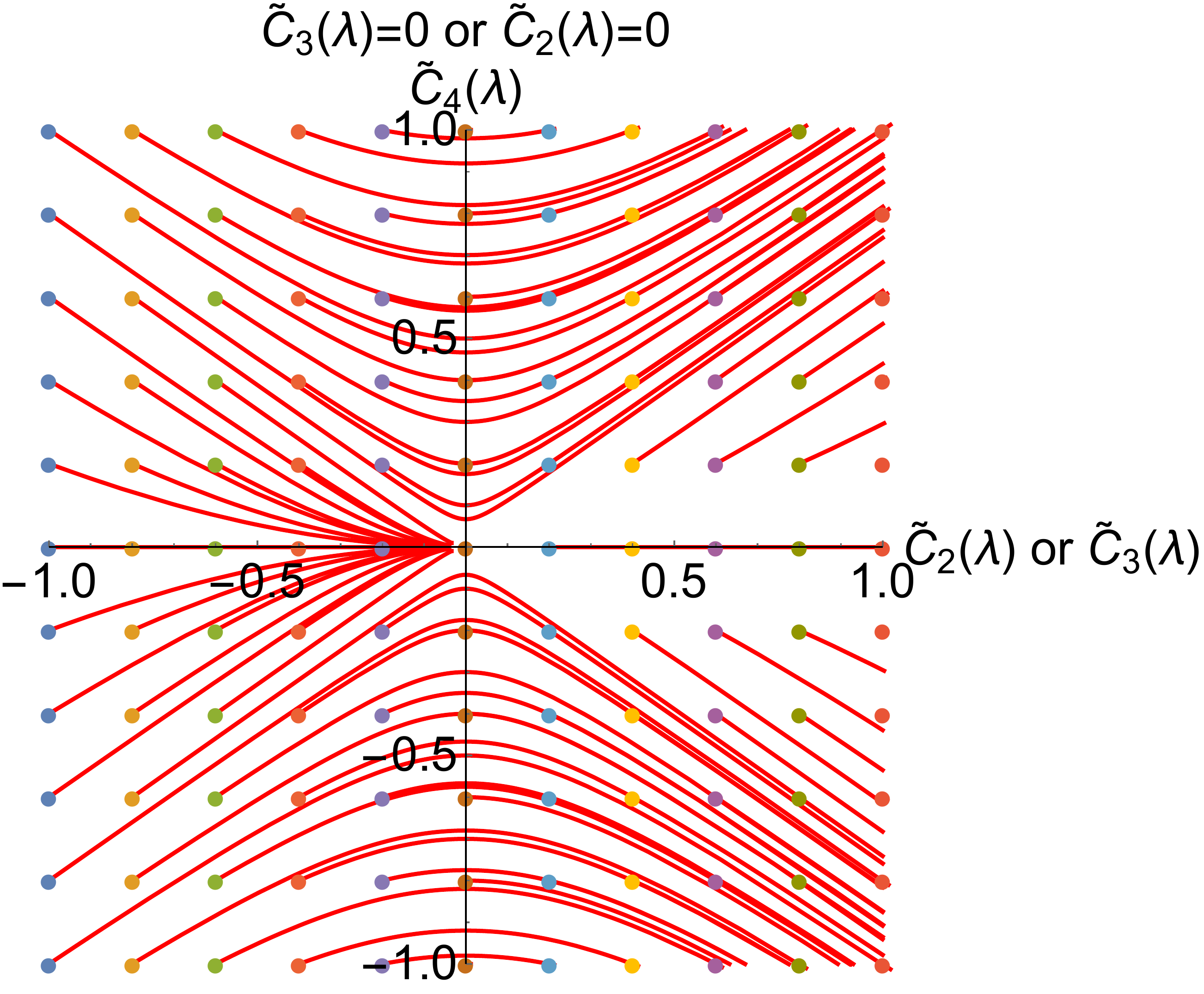}
        \end{center}
      \end{minipage}
    \end{tabular}
\caption{The flow diagram on the $\tilde{C}_{2}(\lambda)$-$\tilde{C}_{4}(\lambda)$ plane for $\tilde{C}_{3}(\lambda)=0$ (or the $\tilde{C}_{3}(\lambda)$-$\tilde{C}_{4}(\lambda)$ plane for $\tilde{C}_{2}(\lambda)=0$). Left: the plot of the vector $\bigl(\tilde{C}_{2}(\lambda)^{2} + 2\tilde{C}_{4}(\lambda)^{2},2\tilde{C}_{2}(\lambda) \tilde{C}_{4}(\lambda)\bigr)$ on the right-hand side of Eq.~\eqref{eq:Sigma_c_Kondo_RG_i0}. The gray region is the region where the effective coupling constants become zero in the low-energy limit and the Kondo effect does not occur. The Kondo effect can occur outside this gray region. Right: the solution $(\tilde{C}_{2}(\lambda),\tilde{C}_{4}(\lambda))$ in Eq.~\eqref{eq:Sigma_c_Kondo_RG_i_sol} is shown by the red lines, where the initial conditions in each line are denoted by the points. The initial points in the gray region in the left converge into zero in the low-energy limit, while the other initial points become divergent.}
\label{fig:190306_Fig4}
\end{center}
\end{figure}

In Fig.~\ref{fig:190306_Fig4}, we plot the region where the Kondo effect occurs.
In the left panel, we show the two-dimensional vector $\bigl(\tilde{C}_{2}(\lambda)^{2} + 2\tilde{C}_{4}(\lambda)^{2},2\tilde{C}_{2}(\lambda) \tilde{C}_{4}(\lambda)\bigr)$, i.e., the right-hand side in Eq.~\eqref{eq:Sigma_c_Kondo_RG_i0}. 
The gray region is the area of $\displaystyle \tilde{C}_{4}(\lambda)<-\tilde{C}_{2}(\lambda)/\sqrt{2}$ and $\displaystyle \tilde{C}_{4}(\lambda)>\tilde{C}_{2}(\lambda)/\sqrt{2}$.
In the right-hand panel, the solution Eq.~\eqref{eq:Sigma_c_Kondo_RG_i_sol} is shown by the streaming red lines.
The initial values of $(\tilde{C}_{2}(\lambda),\tilde{C}_{4}(\lambda))$ are denoted by the points.
When the initial points are in the gray region (the left-hand panel), the effective coupling constants become zero at the end of the low-energy scale, which indicates that the Kondo effect does not occur.
On the other hand, when the initial points are outside the gray region (the left-hand panel), the effective coupling constants become infinity, and accordingly the Kondo effect occurs.
Here the existence of the $C_{4}$ term is important.
In Sec.~\ref{sec:C40}, we showed that the negative value of $\tilde{C}_{2}(\lambda)$ has not led to the Kondo effect, when the $C_{4}$ term is absent ($C_{4}=0$).
However, when the $C_{4}$ term is present ($C_{4}\neq0$), the negative value of $\tilde{C}_{2}(\lambda)$ can produce the Kondo effect as long as $\displaystyle \tilde{C}_{4}(\lambda)>-\tilde{C}_{2}(\lambda)/\sqrt{2}$ or $\displaystyle \tilde{C}_{4}(\lambda)<\tilde{C}_{2}(\lambda)/\sqrt{2}$ is satisfied.
Therefore, we conclude that the nonzero value of $|\tilde{C}_{4}(\lambda)|$ is important to enhance the parameter region of $\tilde{C}_{2}(\lambda)$ to realize the Kondo effect.
The above conclusion applies also to the case for the two-dimensional space $\bigl(\tilde{C}_{3}(\lambda),\tilde{C}_{4}(\lambda)\bigr)$ with $\tilde{C}_{2}(\lambda)=0$.

\subsubsection{Two-dimensional case II}

We consider the solutions in the case of $\tilde{C}_{2}(\lambda)=\tilde{C}_{3}(\lambda)$.
In addition, we suppose a small value of $|\tilde{C}_{4}(\lambda)|$.
For convenience, we introduce a function $\tilde{C}_{23}(\lambda)\equiv\tilde{C}_{2}(\lambda)=\tilde{C}_{3}(\lambda)$, expressing the RG equations for $\tilde{C}_{23}(\lambda)$ and $\tilde{C}_{4}(\lambda)$ as
\begin{eqnarray}
   \frac{\drm}{\drm \lambda} \tilde{C}_{23}(\lambda) &=& \tilde{C}_{23}(\lambda)^{2} + 2\tilde{C}_{4}(\lambda)^{2},
\nonumber \\
   \frac{\drm}{\drm \lambda} \tilde{C}_{4}(\lambda) &=& 4\tilde{C}_{23}(\lambda) \tilde{C}_{4}(\lambda).
\label{eq:Sigma_c_Kondo_RG_ii0}
\end{eqnarray}
By eliminating $\tilde{C}_{23}(\lambda)$ in the two equations, we find the simple equation for $\tilde{C}_{4}(\lambda)$:
\begin{eqnarray}
     \tilde{C}_{4}(\lambda) \frac{\drm^{2}}{\drm \lambda^{2}} \tilde{C}_{4}(\lambda)
   - \frac{5}{4} \biggl( \frac{\drm}{\drm \lambda} \tilde{C}_{4}(\lambda) \biggr)^{2}
   - 8\tilde{C}_{4}(\lambda)^{4}
=0.
\label{eq:Sigma_c_Kondo_RG_ii}
\end{eqnarray}
For a further simplification, we introduce the function $F(\lambda)$ defined by
$\displaystyle \tilde{C}_{4}(\lambda) = 1/\bigl( F(\lambda) \bigr)^{4}$
with $F(\lambda)>0$.
Then, the equation for $F(\lambda)$ reads
\begin{eqnarray}
   \frac{\drm^{2}}{\drm \lambda^{2}}F(\lambda) + \frac{2}{F(\lambda) ^{7}}
=0,
\end{eqnarray}
which looks much simpler than Eq.~\eqref{eq:Sigma_c_Kondo_RG_ii}.
However, it is still difficult in general to find an analytical solution of $F(\lambda)$.
Here we try to find an approximate solution, and for this purpose we restrict our attention to a small value of $|\tilde{C}_{4}(\lambda)|$, i.e., $|\tilde{C}_{4}(\lambda)| \ll 1$ or $F(\lambda) \gg 1$, where the perturbation can be used.
Then, the equation for $F(\lambda)$ is reduced to
$\displaystyle {\drm^{2}F(\lambda)}/{\drm \lambda^{2}} \approx 0$,
and the solution is found to be
$F(\lambda) \simeq c_{1} \lambda + c_{2}$
with the appropriate constants $c_{1}$ and $c_{2}$.
The values of $c_{1}$ and $c_{2}$ should be fixed by the initial condition of $\tilde{C}_{23}(\lambda)$ and $\tilde{C}_{4}(\lambda)$ at $\lambda=0$.
Finally, we obtain the approximate solution
\begin{eqnarray}
   \tilde{C}_{23}(\lambda) &\simeq& \frac{2\tilde{C}_{23}}{2-\tilde{C}_{23}\lambda}, \nonumber \\
   \tilde{C}_{4}(\lambda) &\simeq& \frac{16\tilde{C}_{4}}{\bigl( 2-\tilde{C}_{23}\lambda \bigr)^{4}},
\end{eqnarray}
with $\tilde{C}_{23}=4\rho_{0}C_{2}=-4\rho_{0}C_{3}$ and $\tilde{C}_{4}=4\rho_{0}C_{4}$ as the initial condition.
The perturbative approach involving the above solution requires that
$\tilde{C}_{23}(\lambda)$ and $\tilde{C}_{4}(\lambda)$ should not be divergent, and
the denominators in $\tilde{C}_{23}(\lambda)$ and $\tilde{C}_{4}(\lambda)$ should satisfy $2-\tilde{C}_{23}\lambda>0$ for any $\lambda>0$.
It indicates that the range of the value of $\tilde{C}_{23}(\lambda)$ should be restricted to $\tilde{C}_{23}(\lambda)<0$ as long as the value of $|\tilde{C}_{4}(\lambda)|$ is small ($|\tilde{C}_{4}(\lambda)| \ll 1$).

\begin{figure}[tb]
\begin{center}
    \begin{tabular}{c}
      \begin{minipage}[t]{0.33\hsize}
        \begin{center}
          \raisebox{0em}{\includegraphics[scale=0.25]{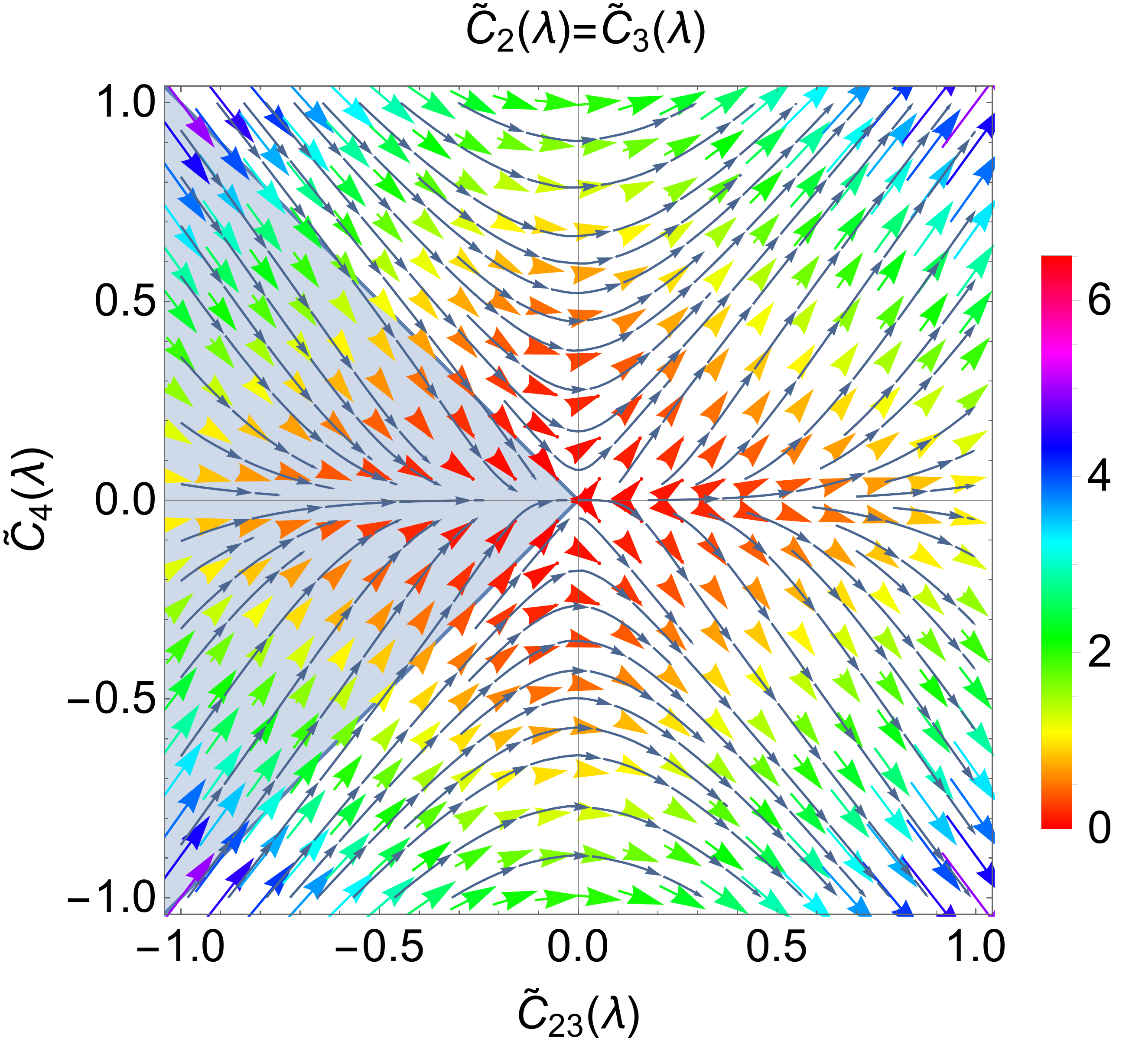}}
        \end{center}
      \end{minipage}
      \hspace{4em}
      \begin{minipage}[t]{0.33\hsize}
      \vspace{-0.6em}
        \begin{center}
          \includegraphics[scale=0.25]{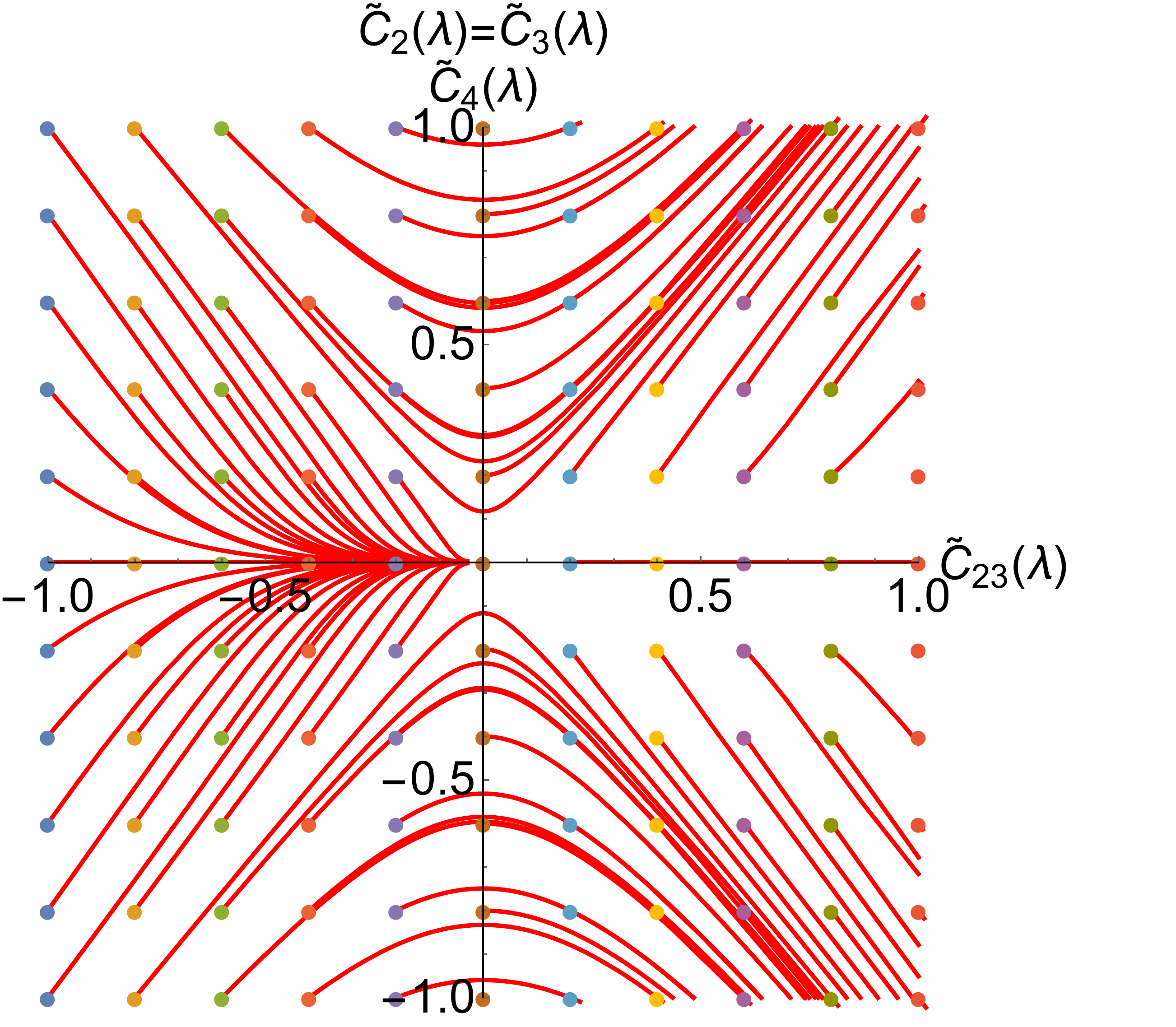}
        \end{center}
      \end{minipage}
    \end{tabular}
\caption{The flow diagram on the $\tilde{C}_{23}(\lambda)$-$\tilde{C}_{4}(\lambda)$ plane. Left: the plot of the vector $\bigl(\tilde{C}_{23}(\lambda)^{2} + 2\tilde{C}_{4}(\lambda)^{2},4\tilde{C}_{23}(\lambda) \tilde{C}_{4}(\lambda)\bigr)$ on the right-hand side of Eq.~\eqref{eq:Sigma_c_Kondo_RG_ii0}. The gray region is the region where the effective coupling constants become zero in the low-energy limit and the Kondo effect does not occur. The Kondo effect can occur outside this gray region. Right: the solution $(\tilde{C}_{23}(\lambda),\tilde{C}_{4}(\lambda))$ in Eq.~\eqref{eq:Sigma_c_Kondo_RG_ii0} is shown by the red lines, where the initial conditions in each line are expressed by the points. The initial points in the gray region in the left converge into zero in the low-energy limit, while the other initial points become divergent.}
\label{fig:190306_Fig5}
\end{center}
\end{figure}

In Fig.~\ref{fig:190306_Fig5}, we plot the two-dimensional vector field $\bigl(\tilde{C}_{23}(\lambda)^{2} + 2\tilde{C}_{4}(\lambda)^{2},4\tilde{C}_{23}(\lambda) \tilde{C}_{4}(\lambda)\bigr)$, i.e., the right-hand side of Eq.~\eqref{eq:Sigma_c_Kondo_RG_ii0}.
We also plot the solutions $\bigl(\tilde{C}_{23}(\lambda),\tilde{C}_{4}(\lambda)\bigr)$ starting from $\lambda=0$ by the streaming lines.
It is shown that the solutions from the initial points with the negative value of $\tilde{C}_{23}(\lambda)$ ($\tilde{C}_{23}(\lambda)<0$) and the small value of $|\tilde{C}_{4}(\lambda)|$ ($|\tilde{C}_{4}(\lambda)|\ll1$) become convergent to zero for $\lambda \rightarrow \infty$.
From the numerical calculation, we find that the initial points in the gray region defined by $\tilde{C}_{23}(\lambda)>\tilde{C}_{4}(\lambda)$ and $\tilde{C}_{23}(\lambda)<-\tilde{C}_{4}(\lambda)$ do not lead to the divergence.
The initial points outside this gray region can lead to the divergence and therefore can produce the Kondo effect.
From the above analysis, we conclude that the nonzero value of $\tilde{C}_{4}(\lambda)$ extends the parameter region of $\bigl(\tilde{C}_{23}(\lambda),\tilde{C}_{4}(\lambda)\bigr)$ for the Kondo effect.

\subsubsection{One-dimensional case ---the $\mathrm{SU}(4)$ limit---}
\label{sec:SU4_limit}

Finally, we consider the one-dimensional case that the parameter $\bigl(\tilde{C}_{2}(\lambda),\tilde{C}_{3}(\lambda),\tilde{C}_{4}(\lambda)\bigr)$ is restricted to the one-dimensional space $\tilde{C}_{2}(\lambda)=\tilde{C}_{3}(\lambda)=\pm\sqrt{2/3}\,\tilde{C}_{4}(\lambda)$.
In this case, we introduce the function $\tilde{C}(\lambda)$ defined by
$\displaystyle \tilde{C}(\lambda)\equiv\tilde{C}_{2}(\lambda)=\tilde{C}_{3}(\lambda)=\pm\sqrt{2/3}\,\tilde{C}_{4}(\lambda)$ for the short notation.
Here, $\mathrm{SU}(2)_{\mathrm{spin}} \times \mathrm{SU}(2)_{\mathrm{isospin}}$ in the Lagrangian \eqref{eq:Lagrangian_Sigmac_LO_4a} happens to be extended to the $\mathrm{SU}(4)$ symmetry according to the implication relation $\mathrm{SU}(2) \times \mathrm{SU}(2) \subset \mathrm{SU}(4)$.
We call this one-dimensional case the $\mathrm{SU}(4)$ limit.
The $\mathrm{SU}(4)$ symmetry is made explicit by introducing the operators of the 15 generators in the $\mathrm{SU}(4)$ symmetry,
$\lambda^{a}/2$ or $\rho^{a}$ ($a=1,2,\dots,15$),
where the operators $\lambda^{a}$ and $\rho^{a}$ are defined by
\begin{eqnarray}
 \lambda^{1} &=& \sigma^{1} \otimes \vec{1}_{2}, \hspace{0.5em}
 \lambda^{2} = \sigma^{2} \otimes \vec{1}_{2}, \hspace{0.5em}
 \lambda^{3} = \sigma^{3} \otimes \vec{1}_{2},
\nonumber \\ 
 \lambda^{4} &=& \vec{1}_{2} \otimes \tau^{1}, \hspace{0.5em}
 \lambda^{5} = \vec{1}_{2} \otimes \tau^{2}, \hspace{0.5em}
 \lambda^{6} = \vec{1}_{2} \otimes \tau^{3},
\nonumber \\ 
 \lambda^{7} &=& \pm \sigma^{1} \otimes \tau^{1}, \hspace{0.5em}
 \lambda^{8} = \pm \sigma^{1} \otimes \tau^{2}, \hspace{0.5em}
 \lambda^{9} = \pm \sigma^{1} \otimes \tau^{3},
\nonumber \\ 
 \lambda^{10} &=& \pm \sigma^{2} \otimes \tau^{1}, \hspace{0.5em}
 \lambda^{11} = \pm \sigma^{2} \otimes \tau^{2}, \hspace{0.5em}
 \lambda^{12} = \pm \sigma^{2} \otimes \tau^{3},
\nonumber \\ 
 \lambda^{13} &=& \pm \sigma^{3} \otimes \tau^{1}, \hspace{0.5em}
 \lambda^{14} = \pm \sigma^{3} \otimes \tau^{2}, \hspace{0.5em}
 \lambda^{15} = \pm \sigma^{3} \otimes \tau^{3},
\label{eq:lambda_def}
\end{eqnarray}
and
\begin{eqnarray}
 \rho^{1} &=& s^{1} \otimes \vec{1}_{2} \otimes \vec{1}_{3}, \hspace{0.5em}
 \rho^{2} = s^{2} \otimes \vec{1}_{2} \otimes \vec{1}_{3}, \hspace{0.5em}
 \rho^{3} = s^{3} \otimes \vec{1}_{2} \otimes \vec{1}_{3},
\nonumber \\ 
 \rho^{4} &=& \vec{1}_{3} \otimes \vec{1}_{2} \otimes t^{1}, \hspace{0.5em}
 \rho^{5} = \vec{1}_{3} \otimes \vec{1}_{2} \otimes t^{2}, \hspace{0.5em}
 \rho^{6} = \vec{1}_{3} \otimes \vec{1}_{2} \otimes t^{3},
\nonumber \\ 
 \rho^{7} &=& \sqrt{\frac{3}{2}} \, s^{1} \otimes \vec{1}_{2} \otimes t^{1}, \hspace{0.5em}
 \rho^{8} = \sqrt{\frac{3}{2}} \, s^{1} \otimes \vec{1}_{2} \otimes t^{2}, \hspace{0.5em}
 \rho^{9} = \sqrt{\frac{3}{2}} \, s^{1} \otimes \vec{1}_{2} \otimes t^{3},
\nonumber \\ 
 \rho^{10} &=& \sqrt{\frac{3}{2}} \, s^{2} \otimes \vec{1}_{2} \otimes t^{1}, \hspace{0.5em}
 \rho^{11} = \sqrt{\frac{3}{2}} \, s^{2} \otimes \vec{1}_{2} \otimes t^{2}, \hspace{0.5em}
 \rho^{12} = \sqrt{\frac{3}{2}} \, s^{2} \otimes \vec{1}_{2} \otimes t^{3},
\nonumber \\ 
 \rho^{13} &=& \sqrt{\frac{3}{2}} \, s^{3} \otimes \vec{1}_{2} \otimes t^{1}, \hspace{0.5em}
 \rho^{14} = \sqrt{\frac{3}{2}} \, s^{3} \otimes \vec{1}_{2} \otimes t^{2}, \hspace{0.5em}
 \rho^{15} = \sqrt{\frac{3}{2}} \, s^{3} \otimes \vec{1}_{2} \otimes t^{3}.
\label{eq:rho_def}
\end{eqnarray}
We keep using the notations $A \otimes B$ and $A \otimes B \otimes C$ which were introduced in Eq.~\eqref{eq:Lagrangian_Sigmac_LO_4a_int}, along with the anti-symmetric tensor $\varepsilon^{ijk}$.
$\lambda^{a}$ and $\rho^{a}$ are normalized as $\mathrm{tr}\,\lambda^{a}\lambda^{b}=4\,\delta^{ab}$ and $\mathrm{tr}\,\rho^{a}\rho^{b}=12\,\delta^{ab}$, respectively.
Adopting the restriction of the parameter $\tilde{C}_{2}(\lambda)=\tilde{C}_{3}(\lambda)=\pm\sqrt{2/3}\,\tilde{C}_{4}(\lambda)$ and the operators $\lambda^{a}$ and $\rho^{a}$,
we rewrite the Lagrangian \eqref{eq:Lagrangian_Sigmac_LO_4} as
\begin{eqnarray}
   {\cal L}_{\mathrm{int}}[\psi,\Psi_{v}^{i}] 
=
C_{1} \, \varphi^{\dag} \varphi\, \bar{\Psi}_{v} \Psi_{v}
-C
\sum_{a=1}^{15}
   \, \varphi^{\dag} \lambda^{a} \varphi
   \, \bar{\Psi}_{v} \rho^{a} \Psi_{v},
\label{eq:Lagrangian_SU4}
\end{eqnarray}
with $C \equiv C_{2}=C_{3}=\pm\sqrt{2/3}\,C_{4}$.
It is easy to prove that Eq.~\eqref{eq:Lagrangian_SU4} is invariant under the $\mathrm{SU}(4)$ symmetry.
We notice that the $\mathrm{SU}(2)_{\mathrm{spin}}$ symmetry and the $\mathrm{SU}(2)_{\mathrm{isospin}}$ symmetry are unified to the $\mathrm{SU}(4)$ symmetry.
Thus, it provides the Kondo effect for a single non-Abelian symmetry.
Regarding the coupling constant $C$ as the effective coupling constant dependent on the energy scale $C(\lambda)$,
the RG equations reads
\begin{eqnarray}
   \frac{\drm}{\drm \lambda} \tilde{C}(\lambda) = 4\tilde{C}(\lambda)^{2},
\end{eqnarray}
with $\tilde{C}(\lambda)\equiv4\rho_{0}C(\lambda)$.
This is indeed obtained by setting $\tilde{C}(\lambda)=\tilde{C}_{2}(\lambda)=\tilde{C}_{3}(\lambda)=\pm\sqrt{2/3}\,\tilde{C}_{4}(\lambda)$ in Eq.~\eqref{eq:RG_Kondo_Sigma_1}.
The solution is given in a simple equation as
\begin{eqnarray}
   \tilde{C}(\lambda) = \frac{\tilde{C}}{1-4\tilde{C}\lambda},
\end{eqnarray}
with $\tilde{C}=4\rho_{0}C$ as the initial condition at $\lambda=0$.
The region of the parameter space for the Kondo effect is limited to $\tilde{C}>0$, i.e., $\tilde{C}_{2}=\tilde{C}_{3}>0$.
From the RG equation, we obtain the Kondo scale $\Lambda_{\mathrm{K}}=\Lambda_{0}e^{-1/(4\tilde{C}_{i})}$ ($i=1,2$) with $\Lambda_{0}$ being the initial energy scale for the RG flow.
We note that the sign of $\tilde{C}_{4}$ is irrelevant to the condition for the Kondo effect, because of the positive and negative signs in $\tilde{C}(\lambda)=\pm\sqrt{2/3}\,\tilde{C}_{4}(\lambda)$.
In the right panel in Fig.~\ref{fig:Fig3}, we plot the line (manifold) which is constrained by $\tilde{C}_{2}(\lambda)=\tilde{C}_{3}(\lambda)=\pm\sqrt{2/3}\,\tilde{C}_{4}(\lambda)$ as the $\mathrm{SU}(4)$ limit.
We observe that the nonzero value of $|\tilde{C}_{4}(\lambda)|$ leads to the Kondo effect.
The relevant sign of $\tilde{C}_{4}(\lambda)$ is either of $\tilde{C}_{4}(\lambda)>0$ for $\tilde{C}_{2}(\lambda)>0$ and $\tilde{C}_{3}(\lambda)>0$ or $\tilde{C}_{4}(\lambda)<0$ for $\tilde{C}_{2}(\lambda)<0$ and $\tilde{C}_{3}(\lambda)<0$, depending on $\tilde{C}(\lambda)=\pm\sqrt{2/3}\,\tilde{C}_{4}(\lambda)$.
Thus, the nonzero value of $|\tilde{C}_{4}|$ is important to bring about the Kondo effect in the SU(4) limit.

\subsection{Flow diagrams in general cases}

\begin{figure}[tb]
\begin{center}
\vspace{0em}
\includegraphics[scale=0.21]{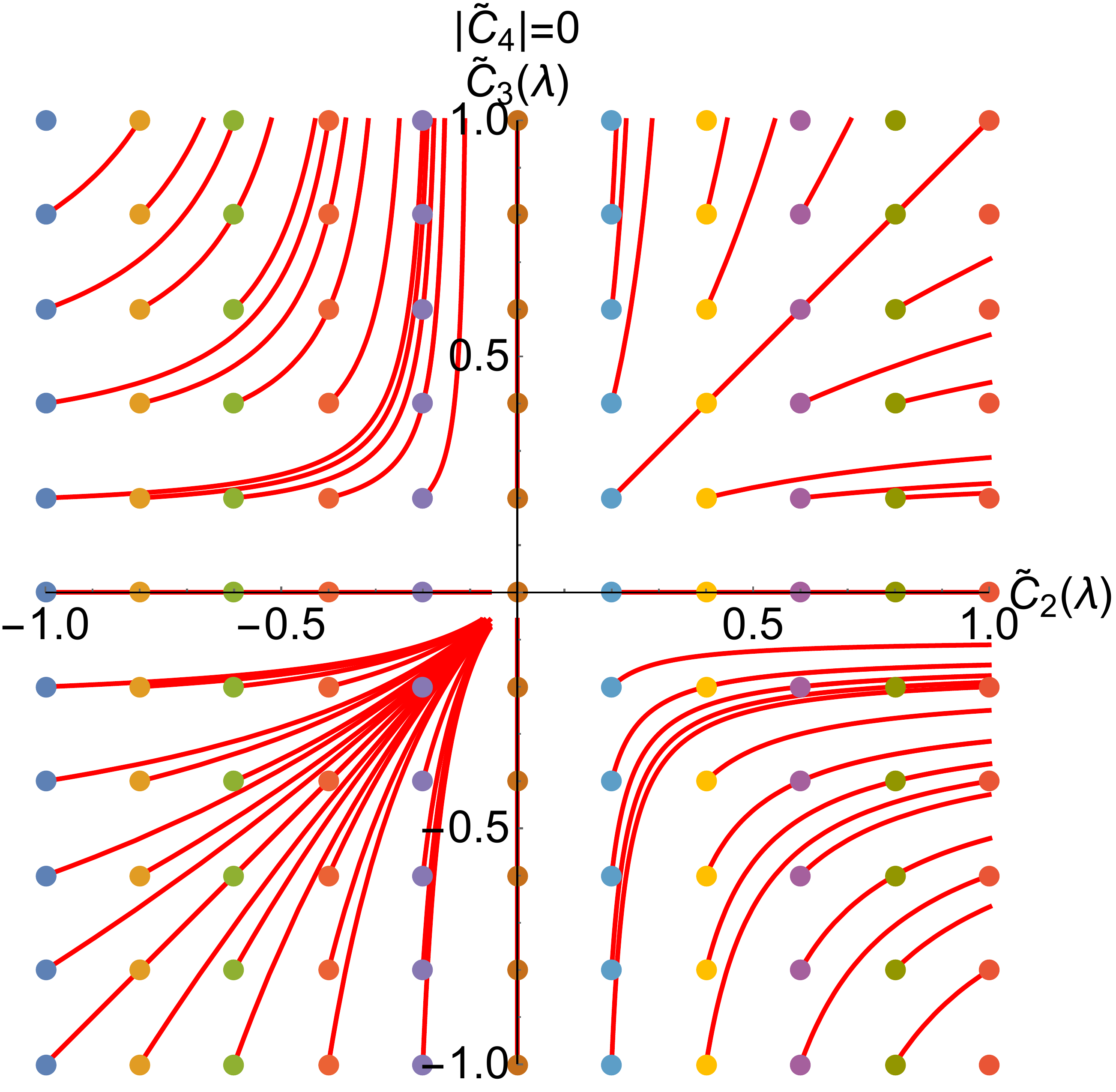}
\hspace{1em}
\includegraphics[scale=0.21]{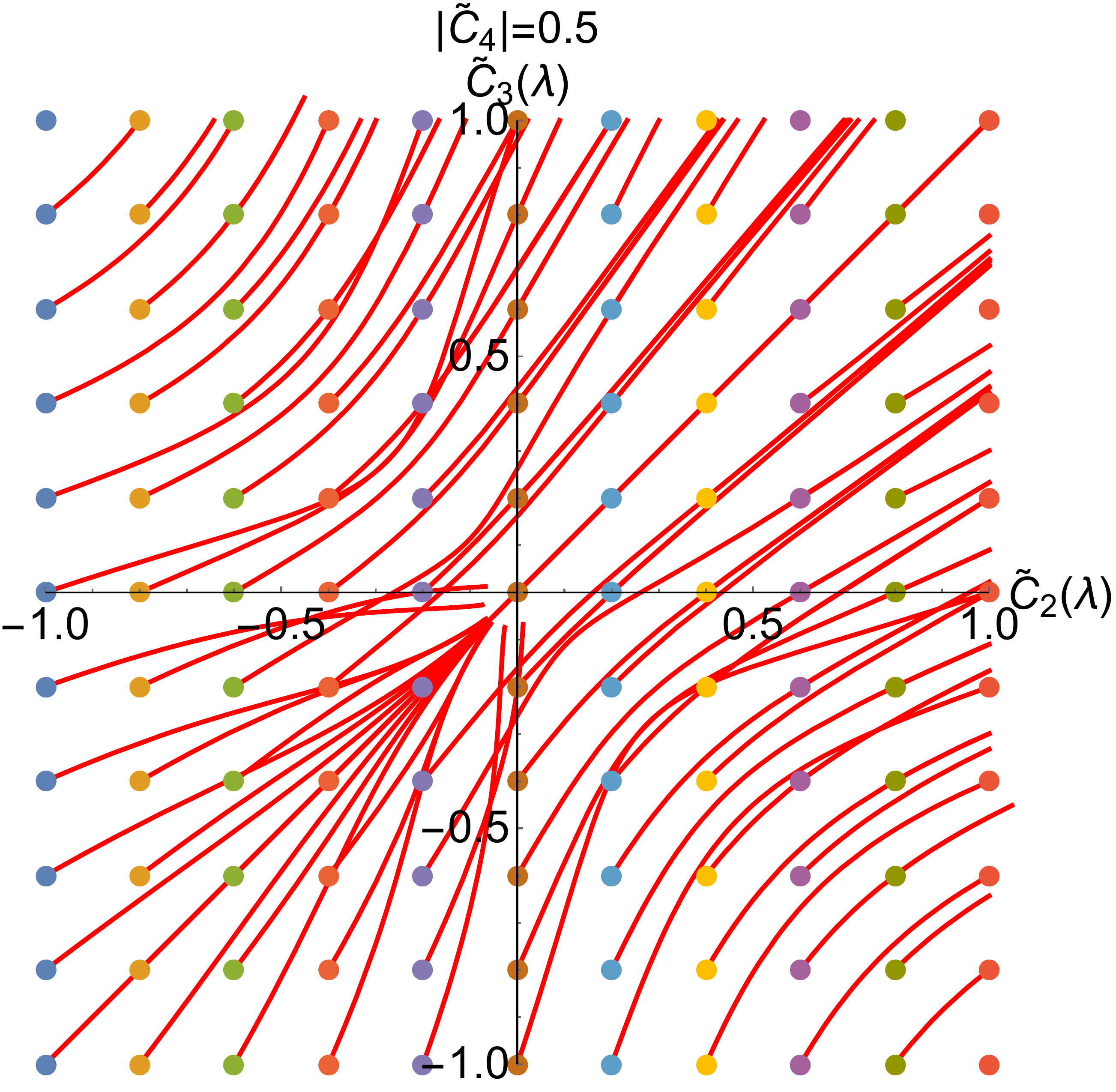}
\hspace{1em}
\includegraphics[scale=0.21]{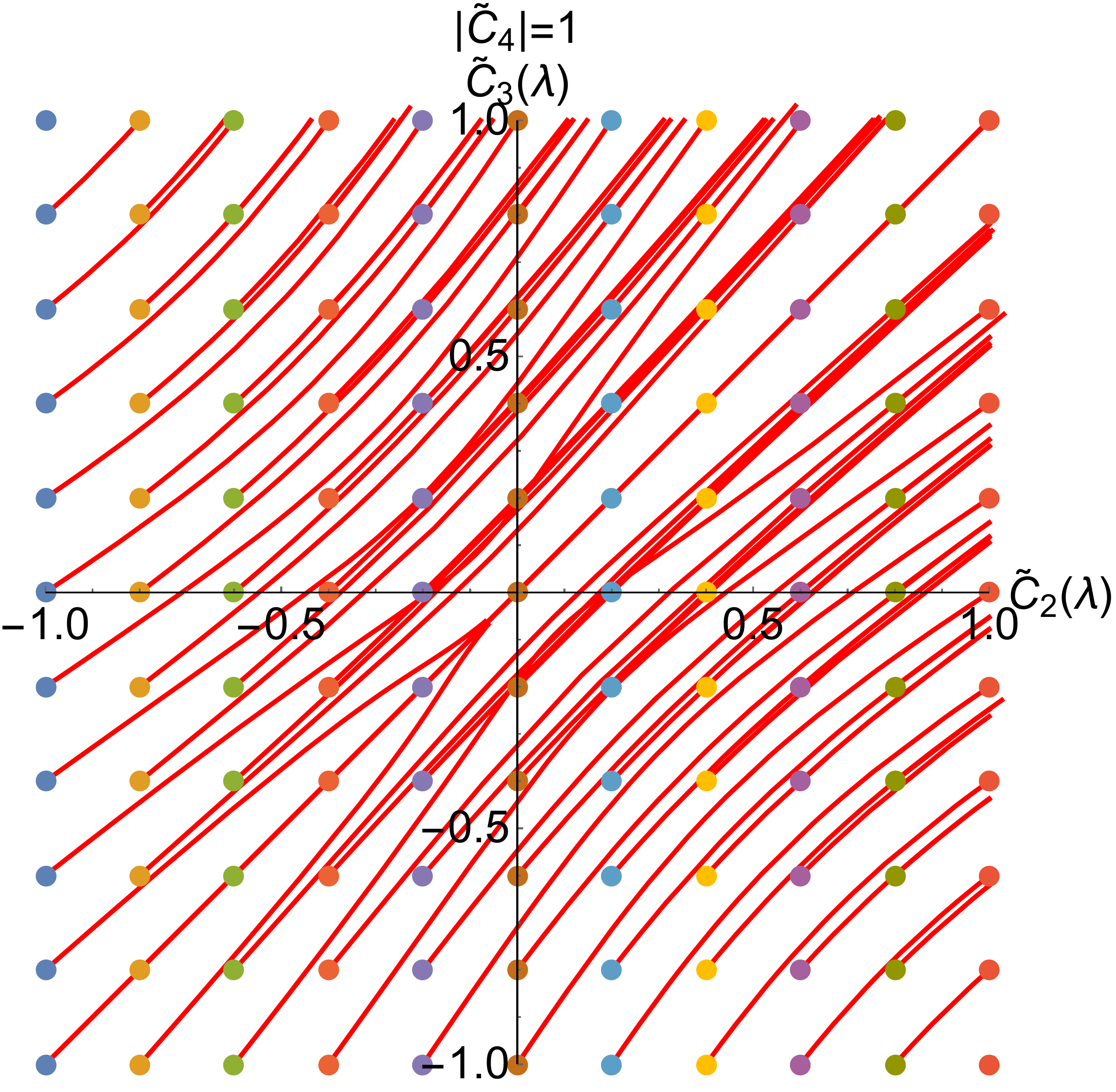}
\vspace{0em}
\caption{The solutions $\tilde{C}_{2}(\lambda)$, $\tilde{C}_{3}(\lambda)$, and $\tilde{C}_{4}(\lambda)$ of the RG equation \eqref{eq:RG_Kondo_Sigma_1} on the two-dimensional $\bigl(\tilde{C}_{2}(\lambda),\tilde{C}_{3}(\lambda)\bigr)$ plane. The initial conditions given by $(\tilde{C}_{2},\tilde{C}_{3})$ at each point and $|\tilde{C}_{4}|=0$, $0.5$, $1$. The solutions are projected to the $\bigl(\tilde{C}_{2}(\lambda),\tilde{C}_{3}(\lambda)\bigr)$ plane when they grow in the three-dimensional $\bigl(\tilde{C}_{2}(\lambda),\tilde{C}_{3}(\lambda),\tilde{C}_{4}(\lambda)\bigr)$ space.}
\label{fig:190306_Fig6}
\end{center}
\end{figure}

In the previous subsections, we highlighted special cases where the nonzero value of $|C_{4}|$, i.e., the spin and isospin-dependent interaction in the Lagrangian \eqref{eq:Lagrangian_Sigmac_LO_4}, extends the parameter region of $\tilde{C}_{2}$, $\tilde{C}_{3}$, and $\tilde{C}_{4}$ and allows the Kondo effect to occur.
As a summary,
we consider the solutions $\bigl(\tilde{C}_{2}(\lambda),\tilde{C}_{3}(\lambda)\bigr)$ which is projected to the two-dimensional surface with a constant value of $\tilde{C}_{4}(\lambda)$.
We suppose the initial conditions of $|\tilde{C}_{4}|=0$, $0.5$, and $1$ for the numerical demonstration.
The results are shown in Fig.~\ref{fig:190306_Fig6}.
For each $C_{4}$, the initial conditions of $\tilde{C}_{2}$ and $\tilde{C}_{3}$ are shown by the dots in the figure.

Under the initial condition of $\tilde{C}_{4}=0$, the Kondo effect occurs for $\tilde{C}_{2}>0$ or $\tilde{C}_{3}>0$ and does not for both $\tilde{C}_{2}<0$ and $\tilde{C}_{3}<0$.
This is confirmed directly in the figure, because the flows in the former are divergent toward large $\tilde{C}_{2}(\lambda)$ and $\tilde{C}_{3}(\lambda)$, while the flows in the latter stop at $\tilde{C}_{2}(\lambda)=\tilde{C}_{3}(\lambda)=0$.
In contrast, if $\tilde{C}_{4}$ has a nonzero value for the initial condition,
the Kondo effect can occur even for both $\tilde{C}_{2}<0$ and $\tilde{C}_{3}<0$.
For example, let us see the initial points of $\tilde{C}_{2}=-1.0$ and $\tilde{C}_{3}=-0.2$ for $|\tilde{C}_{4}|=0.5$ and the initial points of $\tilde{C}_{2}=-1.0$ and $\tilde{C}_{3}=-0.4$ for $|\tilde{C}_{4}|=1$.
Therefore, we understand numerically that the nonzero value of $\tilde{C}_{4}$ helps to extend the region of the parameter space of $\tilde{C}_{2}$ and $\tilde{C}_{3}$ for which the Kondo effect occurs.

Comparison of the Kondo scales allows us to grasp the importance of the $C_{4}$ term, and to do so we
 consider the Kondo scales for the $\mathrm{SU}(2)$ symmetry in $C_{4}=0$ and for the $\mathrm{SU}(4)$ symmetry in $C_{4}\neq0$.
In the former case, assuming $\tilde{C}_{2}=\tilde{C}_{3}$, we have obtained the Kondo scale $\Lambda_{\mathrm{K}}^{\mathrm{SU}(2)}=\Lambda_{0}e^{-1/\tilde{C}_{i}}$ ($i=2,3$) as shown in Sec.~\ref{sec:C40}.
The symmetry is $\mathrm{SU}(2)$, because the $C_{2}$ term and the $C_{3}$ term are completely decoupled.
In the latter case, assuming $\tilde{C}_{2}=\tilde{C}_{3}=\pm\sqrt{2/3}\,\tilde{C}_{4}$,
we have obtained the Kondo scale $\Lambda_{\mathrm{K}}^{\mathrm{SU}(4)}=\Lambda_{0}e^{-1/(4\tilde{C}_{i})}$ ($i=2,3$; $\tilde{C}_{2}=\tilde{C}_{3}$) as shown in Sec.~\ref{sec:SU4_limit}.
As the two Kondo scales are strongly influenced by the exponential factors, their magnitudes are quite different: $\Lambda_{\mathrm{K}}^{\mathrm{SU}(2)} \ll \Lambda_{\mathrm{K}}^{\mathrm{SU}(4)}$.
Therefore, keeping the same coupling constants $C_{2}$ and $C_{3}$,
we find that the $C_{4}$ term, i.e., the mixing term of both spin and isospin, enhances the Kondo scale.
Such enhancement makes the Kondo effect with the nonzero value of $C_{4}$ occur on higher energy scales than the case of $C_{4}=0$.
This conclusion supports the argument that the $C_{4}$ term is important to magnify the Kondo effect for the $\Sigma_{c}$ ($\Sigma_{c}^{\ast}$) baryon in nuclear matter.

\subsection{Estimates of the Kondo scales}

Before ending the discussion about the Kondo effect for the $\Sigma_{c}$ and $\Sigma_{c}^{\ast}$ baryon,
we examine numerically the dependence of the effective coupling constants on the variation of the energy scale.
For example, we can reasonably regard the energy scale $\Lambda$ in the RG equation as the temperature in the system, $T \approx \Lambda$.
As far as we know, however, we have no sufficient data to uniquely determine the coupling constant $C_{i}$ ($i=1,2,3,4$) in the bare Lagrangian~\eqref{eq:Lagrangian_Sigmac_LO_4}.
Thus we have to resort to the semiquantitative approach for restricting the range of the values of $C_{i}$.
As a system similar to the $\Sigma_{c}$ and $\Sigma_{c}^{\ast}$ baryon,
we may consider a $\Lambda_{c}$ baryon with spin 1/2 and isospin 0 whose constituent is $udc$ with a $ud$ diquark.\footnote{The $\Lambda_{c}$ baryon dose not induce the Kondo effect in nuclear matter, because the spin-flip amplitude of the $\Lambda_{c}$ is suppressed by ${\cal O}(1/m_{c})$ with a charm quark mass $m_{c}$.}
As for the interaction between a $\Lambda_{c}$ baryon and a nucleon,
there is an available result from the analysis for the scattering length ($a \approx 0.89$ fm) in the combination of the lattice QCD simulations~\cite{Miyamoto:2017tjs} and the chiral extrapolations~\cite{Haidenbauer:2017dua}.\footnote{See also the related studies of the $\Lambda_{c}$ baryon in nuclear matter~\cite{Yasui:2018sxz,Carames:2018xek,Vidana:2019amb}.}
In Ref.~\cite{Yasui:2018sxz}, based on the HQS formalism in a similar manner as Eq.~\eqref{eq:Lagrangian_Sigmac_LO_4},
the possible combinations of the coupling constants and the three-dimensional momentum cutoff parameter were obtained to reproduce the scattering length $a \approx 0.89$ fm.
They are roughly on the order of 10-16 GeV$^{-2}$ and 0.3-0.5 GeV.
The momentum cutoff parameter would be reasonable because its inverse should be relevant to the spatial size of the hadrons.
As a crude estimate, we may use similar values for $\Sigma_{c}$ and $\Sigma_{c}^{\ast}$ baryons.
As a typical value, we assign $\Lambda_{0} = 0.5$ GeV and $C=C_{i}$ ($i=2,3$) to be around a few GeV$^{-1}$: $C=1$, $2$, $3$ GeV$^{-2}$.
We consider the values smaller than 10-16 GeV$^{-2}$ in order to avoid the overestimates.

\begin{figure}[tb]
\begin{center}
\vspace{0em}
\includegraphics[scale=0.24]{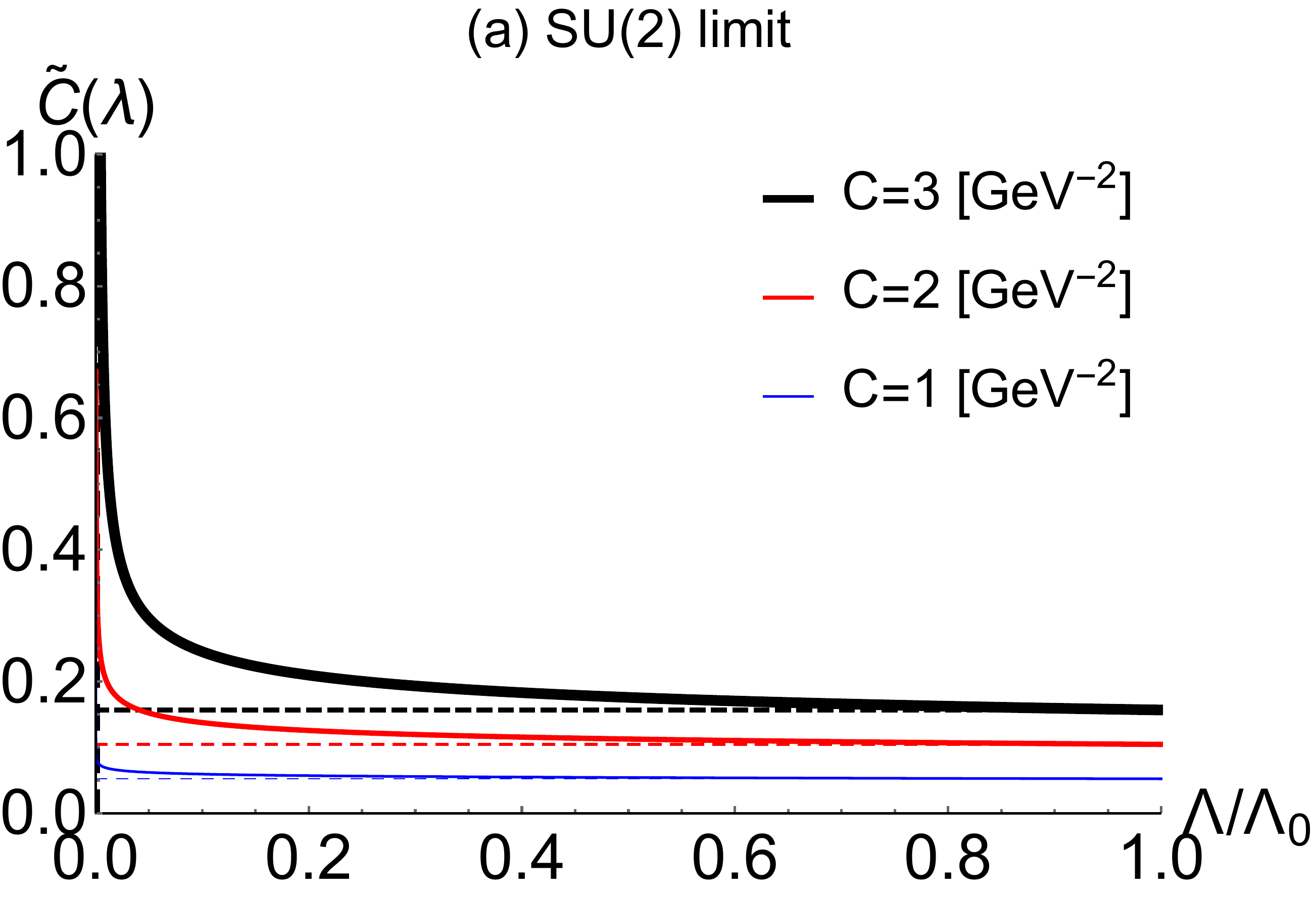}
\hspace{1em}
\includegraphics[scale=0.24]{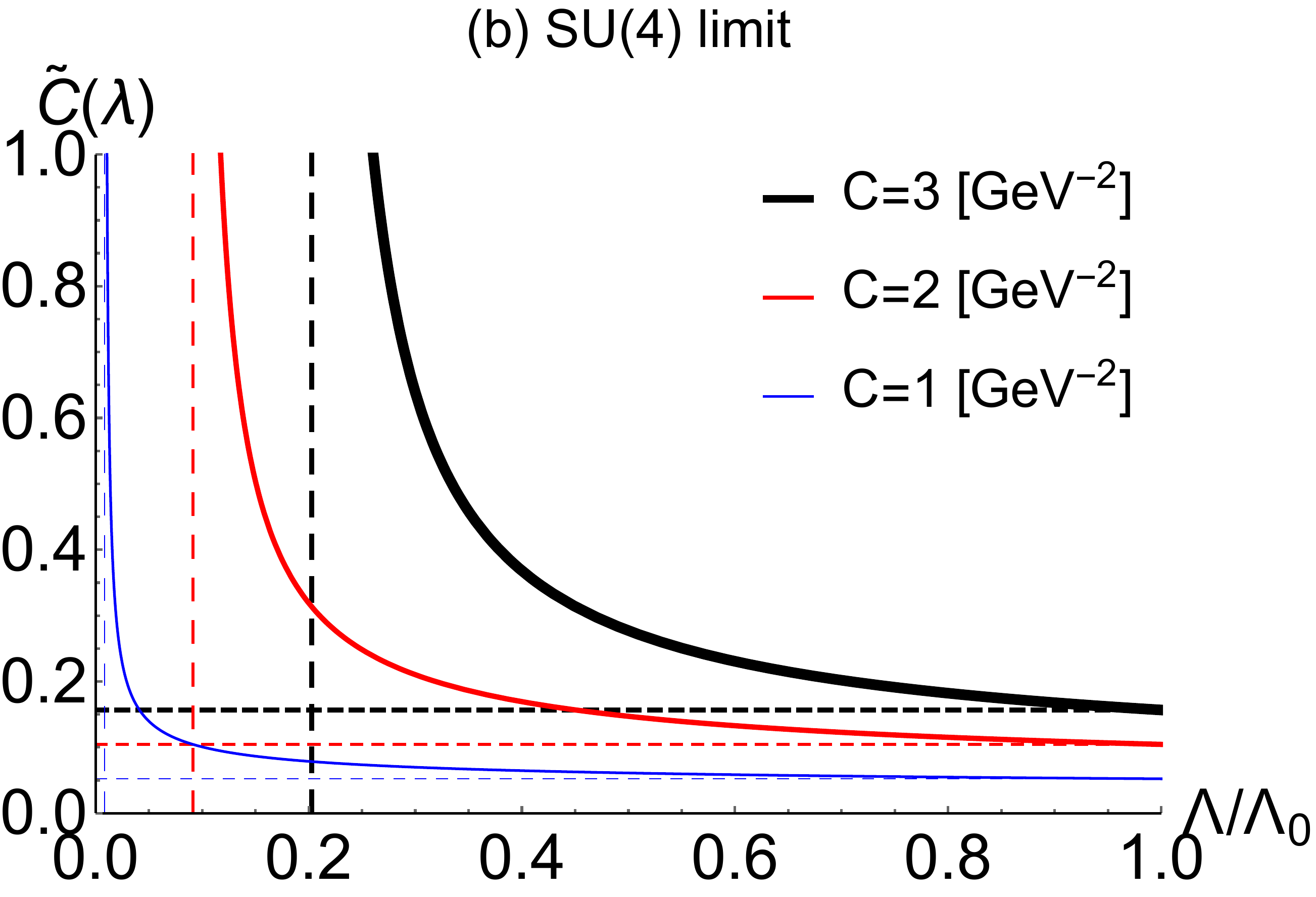}
\vspace{0em}
\caption{The plots of $\tilde{C}(\lambda)=\tilde{C}_{i}(\lambda)$ ($i=2,3$) as functions of $\Lambda/\Lambda_{0}$ for (a) the SU(2) limit and for (b) the SU(4) limit, respectively. The horizontal dashed lines are the values of the coupling constants at the initial condition ($\Lambda=\Lambda_{0}$), and the vertical dashed lines indicate the Kondo energy scales.}
\label{fig:190810_SU2}
\end{center}
\end{figure}

With those values, we estimate the Kondo scale
$\Lambda_{\mathrm{K}}^{\mathrm{SU}(4)}=\Lambda_{0}e^{-1/(16\rho_{0}C_{i})}$ ($i=2,3$; $C_{2}=C_{3}$) with $C_{2}=C_{3}=\pm\sqrt{2/3}\,C_{4}$ in the SU(4) limit (cf.~Sec.~\ref{sec:SU4_limit}).
We remind that $\rho_{0} = m^{3/2}\sqrt{2\mu}/(2\pi^{2})$ is the state-number density at the Fermi surface.
This quantity turns out to be $\rho_{0}=0.0013$ GeV$^{2}$ at $\mu=0.04$ GeV in the normal nuclear matter density.
Then, we obtain $\Lambda_{\mathrm{K}}^{\mathrm{SU}(4)}=0.001$, $0.012$, $0.027$ GeV for $C_{2}=C_{3}=1$, $2$, $3$ GeV$^{-2}$, respectively.
In terms of the temperature in normal nuclear-matter, the values of those Kondo scales are thought to be large, and hence the Kondo effect should occur.
We also consider the Kondo scale $\Lambda_{\mathrm{K}}^{\mathrm{SU}(2)}=\Lambda_{0}e^{-1/(4\rho_{0}C_{i})}$ ($i=2,3$) with $C_{2}=C_{3}$ in the SU(2) limit (cf.~Sec.~\ref{sec:C40}).
In this case, we obtain numerically $\Lambda_{\mathrm{K}}^{\mathrm{SU}(2)}=6\times10^{-10}$, $9 \times 10^{-6}$, $2\times10^{-4}$ GeV for $C_{2}=C_{3}=1$, $2$, $3$ GeV$^{-2}$, respectively.
Thus, the Kondo scale in the SU(2) limit will not be so relevant to the real systems.
We notice that only the difference between the Kondo scale in the SU(4) limit and that in the SU(2) limit lies in the difference of the coefficients in the exponential functions.
In Fig.~\ref{fig:190810_SU2}, we show the plots of $\tilde{C}(\lambda)=\tilde{C}_{i}(\lambda)$ ($i=2,3$) with the bare couplings $C=C_{i}=1,2,3$ GeV$^{-2}$ for the SU(2) limit and the SU(4) limit, where the differences of the Kondo scales are seen clearly.

\section{Revisiting $\bar{D}$ and $\bar{D}^{\ast}$ mesons}
\label{sec:revising_D}

Now we consider other systems where multiple numbers of non-Abelian symmetries exist, and here we focus on the Kondo effect for a $\bar{D}$ ($\bar{D}^{\ast}$) meson in terms of the $\mathrm{SU}(2)$ spin symmetry and the $\mathrm{SU}(2)$ isospin symmetry.
Although there have been many studies on the $\bar{D}N$ ($\bar{D}^{\ast}N$) interaction~\cite{Hofmann:2005sw,Haidenbauer:2007jq,Yasui:2009bz,Yamaguchi:2011xb,Yamaguchi:2011qw,Gamermann:2010zz,Carames:2012bd,Fontoura:2012mz} and the properties of a $\bar{D}$ ($\bar{D}^{\ast}$) meson in nuclear systems~\cite{Tsushima:1998ru,Sibirtsev:1999js,Mishra:2003se,Lutz:2005vx,Tolos:2007vh,Hilger:2008jg,Hilger:2010zb,Mishra:2008cd,Kumar:2010gb,Kumar:2011ff,JimenezTejero:2011fc,GarciaRecio:2011xt,Blaschke:2011yv,Yasui:2012rw,Yasui:2013iga,Yasui:2013vca,Yamaguchi:2014era,Yamaguchi:2013hsa,Suzuki:2015est,Park:2016xrw,Yamaguchi:2016crp,Suenaga:2014dia,Suenaga:2014sga,Suenaga:2015daa,Suenaga:2015wbu,Harada:2016uca,Suenaga:2017deu,Suenaga:2018kta} in the literature, there are only a few studies on the Kondo effect for a $\bar{D}$ ($\bar{D}^{\ast}$) meson.
In the previous studies, only the isospin symmetry was taken in Refs.~\cite{Yasui:2013xr,Yasui:2016ngy}, and only the spin symmetry was taken in Ref.~\cite{Yasui:2016hlz}.
In the present study, we extend their discussions to the case where both of them exist.
We introduce $H_{v}$ defined by
$ H_{v} = \left( \gamma^{\mu}P^{\ast}_{v\mu} + i\gamma_{5}P_{v} \right)\frac{1+v\hspace{-0.5em}/}{2}$
with $P^{\ast}_{v\mu} \sim (q\bar{Q})_{\mathrm{spin}\,1}$ ($\mu=0,1,2,3$) for the vector meson and $P_{v} \sim (q\bar{Q})_{\mathrm{spin}\,0}$ for a pseudoscalar meson.
We note that the asterisk ($\ast$) denotes the vector field, not the complex conjugate.
The vector field satisfies $v^{\mu}P^{\ast}_{v\mu}=0$ and $\bar{H}_{v}= \gamma^{0}H_{v}^{\dag}\gamma^{0}$.
Under the spin and isospin symmetries and the S-wave interaction at the low energies, we write the interaction Lagrangian as follows:
\begin{eqnarray}
 {\cal L}[\psi,H_{v}]
&=&
   {\cal L}_{\mathrm{0}}[\psi,H_{v}] + {\cal L}_{\mathrm{int}}[\psi,H_{v}],
\label{eq:Lagrangian_ND}
\end{eqnarray}
with the kinetic term
\begin{eqnarray}
{\cal L}_{\mathrm{0}}[\psi,H_{v}]
&=&
     \varphi^{\dag}i\frac{\partial}{\partial t}\varphi
  + \varphi^{\dag} \frac{(i\vec{\nabla})^{2}}{2m}\varphi
  + \mathrm{tr} \, \bar{H}_{v} \biggl( -i\frac{\partial}{\partial t} \biggr) H_{v},
\end{eqnarray}
and the interaction term
\begin{eqnarray}
{\cal L}_{\mathrm{int}}[\psi,H_{v}]
=
   \sum_{i=1}^{4} \frac{d^{s}_{i}}{2} \, \bar{\psi} \Gamma_{i} \psi \, \mathrm{tr} \, \bar{H}_{v} \Gamma_{i} H_{v}
+ \sum_{i=1}^{4} \sum_{a=1}^{3} \frac{d^{t}_{i}}{2} \, \bar{\psi} \Gamma_{i}\tau^{a} \psi \, \mathrm{tr} \, \bar{H}_{v} \Gamma_{i} \tau^{a} H_{v}.
\end{eqnarray}
We define the Dirac matrices by 
 $\Gamma_{1}=1$,
 $\Gamma_{2} = \gamma^{\mu}$,
 $\Gamma_{3}=\sigma^{\mu\nu}=(i/2)\left( \gamma^{\mu}\gamma^{\mu}-\gamma^{\nu}\gamma^{\mu} \right)$,
 $\Gamma_{4}=\gamma^{\mu}\gamma_{5}$,
 $\Gamma_{5}=\gamma_{5}$,
and here $\mathrm{tr}$ stands for the trace over the Dirac matrices.
We introduce $d^{s}_{i}$ and $d^{t}_{i}$ ($i=1,2,3,4$) for the coupling constants in each isospin channel (singlet and triplet).
The coefficient $1/2$ is used for later convenience.
In the interaction term, the relativistic field $\psi$ for a nucleon is reduced to the nonrelativistic form as $\psi^{t}=(\varphi,0)^{t}$ in the following discussion.
Then, we rewrite ${\cal L}_{\mathrm{int}}[\psi,H_{v}]$ in terms of $P_{v}$ and $P_{v}^{\ast i}$ as
\begin{eqnarray}
   {\cal L}_{\mathrm{int}}[\psi,H_{v}]
&=& 
    D_{1} \, \varphi^{\dag}\varphi
    \Biggl( \sum_{i,j=1}^{3} \delta^{ij} P_{v}^{\ast i \dag} P^{\ast j}_{v} + P_{v}^{\dag}P_{v} \Biggr)
\nonumber \\ && 
 + iD_{2} \sum_{k=1}^{3} \varphi^{\dag} \sigma^{k} \varphi
    \Biggl(
         \sum_{i,j=1}^{3} \epsilon^{ijk} P_{v}^{\ast i \dag} P^{\ast j}_{v}
       - \Bigl( P_{v}^{\ast k\dag} P_{v} - P_{v}^{\dag}P^{\ast k}_{v} \Bigr)
    \Biggr)
\nonumber \\ && 
 + D_{3} \sum_{a=1}^{3} \varphi^{\dag}\tau^{a}\varphi
    \Biggl( \sum_{i,j=1}^{3} \delta^{ij} P_{v}^{\ast i \dag} \tau^{a} P^{\ast j}_{v} + P_{v}^{\dag}\tau^{a}P_{v} \Biggr)
\nonumber \\ && 
 + iD_{4} \sum_{k=1}^{3} \sum_{a=1}^{3} \varphi^{\dag} \sigma^{k} \tau^{a} \varphi
    \Biggl(
          \sum_{i,j=1}^{3} \epsilon^{ijk} P_{v}^{\ast i \dag} \tau^{a}P^{\ast j}_{v}
        - \Bigl( P_{v}^{\ast k\dag} \tau^{a}P_{v} - P_{v}^{\dag}\tau^{a}P^{\ast k}_{v} \Bigr)
     \Biggr),
\label{eq:DN_int_Lagrangian}
\end{eqnarray}
in the rest frame $v^{\mu}=(1,\vec{0})$,
where we define the new coupling constants by
 $D_{1} = -\bigl(d^{s}_{1}-d^{s}_{2}\bigr)$,
 $D_{2}=2d^{s}_{3}+d^{s}_{4}$,
 $D_{3} = -\bigl(d^{t}_{1}-d^{t}_{2}\bigr)$, and
 $D_{4}=2d^{t}_{3}+d^{t}_{4}$.
Equation~\eqref{eq:DN_int_Lagrangian} is invariant under the flavor symmetry for the light quark and under the spin symmetries for the spin of the light quark and the heavy antiquark.
In terms of the spin symmetry, the transformation of $P_{v}$ and $P_{v}^{\ast i}$ is given by
\begin{eqnarray}
 P_{v} \mapsto P_{v} + \delta P_{v} = P_{v} - \frac{1}{2} \theta^{i} P_{v}^{\ast i}, \hspace{0.5em}
 P_{v}^{\ast i} \mapsto P_{v}^{\ast i} + \delta P_{v}^{\ast i} = P_{v}^{\ast i} + \left( -\frac{1}{2} \epsilon^{ijk} \theta^{j} P_{v}^{\ast k} + \frac{1}{2} \theta^{i} P_{v} \right),
\label{eq:HQS_transformation}
\end{eqnarray}
for the small rotation angle $\theta^{i}$ ($i=1,2,3$).

For the coupling constants in the Lagrangian \eqref{eq:DN_int_Lagrangian}, we consider the effective coupling constants $D_{a}(\lambda)$ ($a=1,2,3,4$) which follows the RG equation, as we have considered for a $\Sigma_{c}$ ($\Sigma_{c}^{\ast}$) baryon in Sec.~\ref{sec:RG_eq_Sigmac}.
Referring to the similar diagram in Fig.~\ref{fig:Fig_190124} and using the momentum integrals \eqref{eq:momentum_integral_1} and \eqref{eq:momentum_integral_2} as well as the identities \eqref{eq:lambda4_Fierz},
we obtain the RG equations
\begin{eqnarray}
   \frac{\drm}{\drm \lambda} \tilde{D}_{1}(\lambda)
&=& 0,
\nonumber \\
   \frac{\drm}{\drm \lambda} \tilde{D}_{2}(\lambda)
&=& 2 \tilde{D}_{2}(\lambda)^{2} + 6 \tilde{D}_{4}(\lambda)^{2},
\nonumber \\
   \frac{\drm}{\drm \lambda} \tilde{D}_{3}(\lambda)
&=& 2 \tilde{D}_{3}(\lambda)^{2} + 6 \tilde{D}_{4}(\lambda)^{2},
\nonumber \\
   \frac{\drm}{\drm \lambda} \tilde{D}_{4}(\lambda)
&=& 4 \Bigl( \tilde{D}_{2}(\lambda) + \tilde{D}_{3}(\lambda) \Bigr) \tilde{D}_{4}(\lambda),
\label{eq:RG_Kondo_D_1}
\end{eqnarray}
with $\lambda = -\ln\bigl(\Lambda/\Lambda_{0}\bigr)$,
where we define the dimensionless quantities by
\begin{eqnarray}
   \tilde{D}_{1}(\lambda) \equiv -\rho_{0}\,D_{1}(\lambda), \hspace{0.5em}
   \tilde{D}_{2}(\lambda) \equiv -\rho_{0}\,D_{2}(\lambda), \hspace{0.5em}
   \tilde{D}_{3}(\lambda) \equiv -\rho_{0}\,D_{3}(\lambda), \hspace{0.5em}
   \tilde{D}_{4}(\lambda) \equiv -\rho_{0}\,D_{4}(\lambda).
\end{eqnarray}
Here, the minus sign is put by convention. 
The RG equation \eqref{eq:RG_Kondo_D_1} is essentially the same as the RG equation for a $\Sigma_{c}$ ($\Sigma_{c}^{\ast}$) baryon, Eq.~\eqref{eq:RG1} or Eq.~\eqref{eq:RG_Kondo_Sigma_1}.
Thus, we obtain the similar behavior for the Kondo effect which indicates the importance of the spin and isospin-dependent term with $D_{4}$.

As a simple case, we consider the $\mathrm{SU}(4)$ limit by setting $D(\lambda) \equiv D_{2}(\lambda)=D_{3}(\lambda)=\pm D_{4}(\lambda)$.
The $\mathrm{SU}(4)$ symmetry is a large group which includes the $\mathrm{SU}(2)_{\mathrm{spin}}$ symmetry and the $\mathrm{SU}(2)_{\mathrm{isospin}}$ symmetry as its subgroups.
In this limit, the RG equation of $\tilde{D}_{i}(\lambda)$ ($i=2,3,4$) is reduced to
\begin{eqnarray}
   \frac{\drm}{\drm \lambda} \tilde{D}(\lambda)
= 8 \tilde{D}(\lambda)^{2},
\label{eq:RG_Kondo_D_SU4}
\end{eqnarray}
with $\tilde{D}(\lambda)\equiv\tilde{D}_{2}(\lambda)=\tilde{D}_{3}(\lambda)=\pm\tilde{D}_{4}(\lambda)$, and we obtain the analytical solution
\begin{eqnarray}
   \tilde{D}(\lambda) = \frac{\tilde{D}}{1-8\tilde{D}\lambda},
\end{eqnarray}
with $\tilde{D}=\tilde{D}(0)$ as the initial condition.
In the low-energy scale (a large value of $\lambda \gg 1$), this solution indicates the divergence
 at the Kondo scale, $\Lambda_{\mathrm{K}}=\Lambda_{0} e^{1/(8\rho_{0}D)}$ for $D<0$, while it leads to the convergence to zero for $D>0$.
 We denote $D \equiv D_{2}=D_{3}=\pm D_{4}$ for the (bare) coupling constant $D_{i}$ ($i=2,3,4$) in the Lagrangian \eqref{eq:DN_int_Lagrangian}.

\section{Discussion: meson-baryon mapping induced by the Kondo effect}
\label{sec:K_mapping}

So far we have demonstrated that the presence of the spin and isospin-exchange term magnifies the Kondo effect, i.e., the increase of the Kondo scale $\Lambda_{\mathrm{K}}$.
This has been based on the perturbative analysis as we have relied on the RG equation.
On the energy scale near or lower than the Kondo scale, however, the perturbative approach is no longer useful due to the enhanced coupling strength,
and hence the nonperturbative approach should be adopted.
So far there have been several nonperturbative analyses
such as the numerical renormalization method~\cite{Wilson:1974mb}, the Bethe ansatz~\cite{PhysRevLett.45.379,Weigmann,RevModPhys.55.331}, the boundary conformal field theory~\cite{Affleck:1990zd,Affleck:1991tk,Affleck:1990by,Affleck:1990iv,Affleck:1992ng,Ludwig:1994nf,Affleck:1995ge}, the bosonization method~\cite{FRADKIN1989710,Ye:1996dj,PhysRevB.9.2911,PhysRevLett.81.196,PhysRevB.61.6918}, the mean-field approximation (the large $N$ limit)~\cite{Takano:1966,Yoshimori:1970,Lacroix:1979,PhysRevB.28.5255,ReadNewns1983,PhysRevLett.57.877,PhysRevB.35.3394,PhysRevB.35.5072,PhysRevLett.79.4665,PhysRevB.58.3794,PhysRevLett.85.1048,Eto:2001,PhysRevB.75.132407,Yanagisawa:2015conf,Yanagisawa:2015}, and so on.
One of the present authors has conducted the analysis based on the mean-field approximation for $D_{s}^{-}$ and $D_{s}^{\ast-}$ mesons in nuclear matter~\cite{Yasui:2016hlz} and for a $\bar{D}$ meson in an atomic nucleus~\cite{Yasui:2016ngy}.
It is still open to question how we should systematically analyze the nonperturbative properties of the Kondo effect on the low-energy scale for both $\Sigma_{c}$ and $\Sigma_{c}^{\ast}$ baryons as well as for both $\bar{D}$ and $\bar{D}^{\ast}$ mesons, where the spin symmetry and the isospin symmetry should be taken into account simultaneously.
In the following, we discuss the expected nonperturbative properties for those systems in a qualitative manner.

It is known that one of the interesting low-energy properties in the Kondo effect is the formation of the singlet pairing in the ground state~\cite{Hewson,Yosida,Yamada}.
Here the singlet pairing indicates the bound state where an itinerant fermion is bound to an impurity particle and the total spin of the bound state is singlet.
In other words, this is the dressed state surrounded by of particles and holes around the impurity site (exact screening).
The dressed state is also known as the Kondo cloud.
The singlet pairing was studied for $D_{s}^{-}$ and $D_{s}^{\ast-}$ mesons in nuclear matter~\cite{Yasui:2016hlz} and for a $\bar{D}$ meson in an atomic nucleus~\cite{Yasui:2016ngy}.
It is also possible that the singlet pairing exists for the $\bar{D}$ and $\bar{D}^{\ast}$ mesons.
In such a situation, the singlet pairing should be composed of a nucleon ($N$) and a light quark ($q=u,d$) in the $\bar{D}$ ($\bar{D}^{\ast}$) meson, i.e., the composite state ($Nq$) with spin $0$ and isospin 0 as the Kondo cloud.
Accordingly, the $\bar{D}$ ($\bar{D}^{\ast}$) meson in nuclear matter should behave as the composite state ($NqQ$), which has the same spin and isospin as a $\Lambda_{c}$ baryon.

For a $D_{s}^{-}$ ($D_{s}^{\ast -}$) meson, 
the singlet pairing as the Kondo cloud is composed of a nucleon ($N$) and the $s$ quark inside the $D_{s}^{-}$ ($D_{s}^{\ast -}$) meson, i.e., the composite state ($Ns$) with spin 0 and isospin 1/2.
In fact, the singlet condensate composed of a nucleon and a $D_{s}^{-}$ ($D_{s}^{\ast -}$) meson was studied in the mean-field approximation~\cite{Yasui:2016hlz}.
Thus, the $D_{s}^{-}$ ($D_{s}^{\ast -}$) meson in nuclear matter should behave like a $\Xi_{c}$ baryon.

In contrast,
the $\Sigma_{c}$ ($\Sigma_{c}^{\ast}$) baryon cannot have the singlet pairing.
In fact it is known that the singlet pairing is not formed when the dimensions of the representations (fundamental, adjoint, etc.) in $\mathrm{SU}(N)$ are different in the itinerant fermion and the impurity particle.
Let us consider the itinerant fermion with spin $1/2$ and the impurity particle with spin $S$.
We observe that, for $S>1/2$, the spin of the impurity particle cannot be screened by the spin of one itinerant fermion, and that there remains an unscreened spin $S^{\ast}=S-1/2$ for the impurity site.
This is called the underscreening Kondo effect~\cite{nozieres:jpa-00209235}.
A similar situation arises for a $\Sigma_{c}$ ($\Sigma_{c}^{\ast}$) baryon in nuclear matter.
That is, the spin $S=1$ and the isospin $I=1$ of the diquark ($qq$) in the $\Sigma_{c}$ ($\Sigma_{c}^{\ast}$) baryon would lead to the unscreened Kondo effect, 
making the $Nqq$ state with the spin $S^{\ast}=1/2$ and the isospin $I^{\ast}=1/2$ as the dressed state by particles and holes.
Furthermore, we argue that it would lead to the composite state of $NqqQ$ with spin $0$ or $1$ and isospin $1/2$, i.e., the same spin and isospin as the $q\bar{Q}$ meson such as a $\bar{D}$ and $\bar{D}^{\ast}$ meson.
Therefore, it is thought that
 the $\Sigma_{c}$ ($\Sigma_{c}^{\ast}$) meson in nuclear matter should behave as the composite state ($NqqQ$), which has the same spin and isospin as a $\bar{D}$ ($\bar{D}^{\ast}$) meson.

The above consideration helps us introduce the concept of the ``meson-baryon mapping" induced by the Kondo effect.
As we have discussed, 
a $\bar{D}$ ($\bar{D}^{\ast}$) meson or a $D_{s}^{-}$ ($D_{s}^{\ast-}$) meson in nuclear matter can be regarded as a $\Lambda_{c}$ baryon or a $\Xi_{c}$ baryon,
and a $\Sigma_{c}$ ($\Sigma_{c}^{\ast}$) baryon in nuclear matter can be regarded as a $\bar{D}$ ($\bar{D}^{\ast}$) meson (Table~\ref{table:Kondo_mapping}).
Thus, the heavy meson is ``baryonized" and the heavy baryon is ``mesonized" due to the Kondo effect.
Such a meson-baryon mapping may cast new light on the properties and the dynamics of heavy hadrons in nuclear matter.
We comment that the simple correspondence between the composite state ($NqQ$ or $NqqQ$) and the hadronlike state ($\Lambda_{c}$-like or $\bar{D}$($\bar{D}^{\ast}$)-like) holds only when both spin and isospin are subject to the Kondo effect.
When only spin (isospin) is subject to the Kondo effect and isospin (spin) is not, there should arise an additional degeneracy by isospin (spin) leading to the hadron-like state whose quantum number is not realized in vacuum.
A more detailed investigation of these things must await another occasion.

\begin{table}[tb]
\begin{center}
\begin{tabular}{|c|c|c|c|}
\hline
  heavy hadron & dressed state (mapped) & screening type & Ref. \\
\hline \hline
   $\bar{D}$, $\bar{D}^{\ast}$ meson & $\Lambda_{c}$ baryon-like & exact screening & --- \\
\hline
   $D_{s}^{-}$, $D_{s}^{\ast-}$ meson & $\Xi_{c}$ baryon-like & exact screening & \cite{Yasui:2016hlz} \\
\hline
   $\Sigma_{c}$, $\Sigma_{c}^{\ast}$ baryon & $\bar{D}$, $\bar{D}^{\ast}$ meson-like & underscreening & --- \\
\hline
\end{tabular}
\end{center}
\caption{The meson-baryon mapping induced by the Kondo effect. See the text for explanation.
}
\label{table:Kondo_mapping}
\end{table}%

\section{Conclusion}

We have studied the Kondo effect for a $\Sigma_{c}$ ($\Sigma_{c}^{\ast}$) baryon in nuclear matter.
By virtue of the $\mathrm{SU}(2)_{\mathrm{spin}} \times \mathrm{SU}(2)_{\mathrm{isospin}}$ symmetry, the HQS symmetry, and the S-wave interaction, we have provided the spin-exchange (or spin-nonexchange) and isospin-exchange (or isospin-nonexchange) interactions between the $\Sigma_{c}$ ($\Sigma_{c}^{\ast}$) baryon and the nucleon.
By adopting the RG equation at one-loop order, we have found that the coexistence of the spin exchange and the isospin exchange magnifies the Kondo effect. 
We have extensively investigated the RG equation for several cases in terms of the coupling constants, including the $\mathrm{SU}(4)$ limit case.
We have also conducted the analysis for the $\bar{D}$ ($\bar{D}^{\ast}$) meson with the $\mathrm{SU}(2)_{\mathrm{spin}} \times \mathrm{SU}(2)_{\mathrm{isospin}}$ symmetry, and have shown the solution in the $\mathrm{SU}(4)$-limit.
In addition, we have ventured to develop the concept of the ``meson-baryon mapping" for the $\Sigma_{c}$ ($\Sigma_{c}^{\ast}$) baryon, the $\bar{D}$ ($\bar{D}^{\ast}$) meson, and the $D_{s}^{-}$ ($D_{s}^{\ast-}$) meson in the Kondo effect.
It is straightforward to apply the mapping to other heavy hadrons when the light component in the heavy hadron has the spin interaction with a nucleon which flips the spin and/or the isospin.

Also, we mention that several issues are left unanswered: the corrections at ${\cal O}(1/m_{Q})$ (beyond the heavy-quark mass limit); applying the $\Sigma_{c}N$ ($\Sigma_{c}^{\ast}N$) interaction to many-body problems~\cite{Maeda:2015hxa,Maeda:2018xcl}; discussing the ``meson-baryon mapping" within the nonperturbative framework; the production mechanisms of the heavy hadrons in atomic nuclei; applications to atomic nuclei.
The existence of the Kondo effect will be verified experimentally through the measurement of the transport coefficients and the change of excitation spectra of atomic nuclei. In the literature, the modifications of the transport coefficients were studied for the Kondo effect in the quark matter~\cite{Yasui:2017bey}, and the excitation spectra for atomic nuclei were studied in a simple model~\cite{Yasui:2016ngy}. Those discussions can be applied to the $\Sigma_{c}$ and $\Sigma_{c}^{\ast}$ baryon (the $\bar{D}$ and $\bar{D}^{\ast}$ meson) in order to prove the existence of the Kondo effect. In order to resolve those problems, it will be important to get precise information of the $\Sigma_{c}N$ ($\Sigma_{c}^{\ast}N$) interactions (see, e.g., Ref.~\cite{Garcilazo:2019ryw} for a recent work).
As a more advanced topic, the continuity of the Kondo effect between the hadronic phase and the quark phase (see the discussions in Ref.~\cite{Hattori:2015hka}).
The continuity, which was proposed for the color-flavor locked color superconductivity in Refs.~\cite{Schafer:1998ef,Alford:1999pa}, is now studied intensively in view of topological objects~\cite{Cipriani:2012hr,Alford:2018mqj,Chatterjee:2018nxe,Cherman:2018jir,Hirono:2018fjr,Hidaka:2019jtv}.
It is worthwhile to study how the Kondo cloud changes from the hadronic matter to the quark matter.
Simulations of the Kondo effect with SU(3) symmetry in cold atomic gases are also important~\cite{Nishida:2013kga}.
It remains unclear as to how the Kondo effect with SU(4) symmetry for a $\Sigma_{c}$ ($\Sigma_{c}^{\ast}$) baryon is related to the Kondo effect with SU(4) symmetry in condensed matter systems, such as quantum dots, which has been studied theoretically~\cite{PhysRevLett.90.026602,doi:10.1143/JPSJ.74.95,PhysRevB.72.085303,SATO2005652,PhysRevB.83.155310,doi:10.7566/JPSJ.85.063702,PhysRevLett.118.196803,PhysRevB.95.245133} and experimentally~\cite{2005Natur.434..484J,PhysRevLett.109.086602,PhysRevB.91.155435,PhysRevLett.121.247703}.
Those issues need to be addressed in future work.

\section*{Acknowledgment}

This work was supported by JSPS Grant-in-Aid for Scientific Research (KAKENHI Grant No.~17K05435), and by the Ministry of Education, Culture, Sports, Science (MEXT)-Supported Program for the Strategic Research Foundation at Private Universities ``Topological Science" (Grant No. S1511006).

\appendix

\section{Spin and charge basis of $\Psi_{v}^{\mu}$}
\label{sec:spin_charge}

We consider the interaction term for the spin-$1/2$ field (a $\Sigma_{c}$ baryon) and the spin-$3/2$ field  (a $\Sigma_{c}^{\ast}$ baryon).
In the rest frame, from Eqs.~\eqref{eq:Psi_2} and \eqref{eq:Psi_4}, we utilize the expression
\begin{eqnarray}
   \Psi_{v1/2}=\frac{1}{\sqrt{3}}\sum_{i}\sigma^{i}\Psi_{v}^{i},
\end{eqnarray}
and
\begin{eqnarray}
   \Psi_{v3/2}^{i}=\sum_{j}\biggl( \delta^{ij} - \frac{1}{3}\sigma^{i}\sigma^{j} \biggr)\Psi_{v}^{j},
\end{eqnarray}
 for the spin-$1/2$ and the spin-$3/2$ fields.\footnote{We take the summation over the indices $i,j=1,2,3$ when they are repeated. Notice the constraint condition $\sigma^{i}\Psi_{v}^{i}=0$.}
Then, we rewrite the interaction term in Eq.~\eqref{eq:Lagrangian_Sigmac_LO_4a_int} as
\begin{eqnarray}
 {\cal L}_{\mathrm{int}}[\psi,\Psi_{v}^{i}] 
&=&
 C_{1} \varphi^{\dag} \varphi \Bigl( \Psi_{v3/2}^{i\dag}\Psi_{v3/2}^{i} + \Psi_{v1/2}^{\dag}\Psi_{v1/2} \Bigr)
\nonumber \\ && 
+
C_{2}
\varphi^{\dag} \sigma^{k} \varphi 
\biggl(
     i\varepsilon^{ijk} \Psi_{v3/2}^{i\dag} \Psi_{v3/2}^{j}
   - \frac{1}{\sqrt{3}} \Psi_{v1/2}^{\dag} \Psi_{v3/2}^{k}
   - \frac{1}{\sqrt{3}} \Psi_{v3/2}^{k\dag} \Psi_{v1/2}
   - \frac{2}{3} \Psi_{v1/2}^{\dag} \sigma^{k} \Psi_{v1/2}
\biggr)
\nonumber \\ && 
+
C_{3} \varphi^{\dag} \tau^{d} \varphi \Bigl( \Psi_{v3/2}^{i\dag} t^{d} \Psi_{v3/2}^{i} + \Psi_{v1/2}^{\dag} t^{d} \Psi_{v1/2} \Bigr)
\nonumber \\ && 
+
C_{4}
\varphi^{\dag} \sigma^{k} \tau^{d} \varphi 
\biggl(
     i\varepsilon^{ijk} \Psi_{v3/2}^{i\dag} t^{d} \Psi_{v3/2}^{j}
   - \frac{1}{\sqrt{3}} \Psi_{v1/2}^{\dag} t^{d} \Psi_{v3/2}^{k}
   - \frac{1}{\sqrt{3}} \Psi_{v3/2}^{k\dag} t^{d} \Psi_{v1/2}
   - \frac{2}{3} \Psi_{v1/2}^{\dag} \sigma^{k} t^{d} \Psi_{v1/2}
\biggr).
\nonumber \\
\end{eqnarray}
Here we mention that the $\Sigma_{c}$ baryon and the $\Sigma_{c}^{\ast}$ baryon can be swapped with each other by the HQS symmetry ($\Sigma_{c} \leftrightarrow \Sigma_{c}^{\ast}$).
In the HQS symmetry,
the heavy quark changes as $u_{v} \rightarrow e^{i\vec{\sigma} \cdot\vec{\theta}/2}u_{v} \approx \bigl(1+i\vec{\sigma}\!\cdot\!\vec{\theta}/2\bigr) u_{v}$ with $\vec{\theta}=(\theta^{1},\theta^{2},\theta^{3})$ for the small $\theta^{i}$ ($i=1,2,3$).
This transformation leads to the change of the fields of $\Sigma_{c}$ and $\Sigma_{c}^{\ast}$ baryons: $\Psi_{v}^{i}$: $\Psi_{v}^{i} \rightarrow e^{i\vec{\sigma} \cdot\vec{\theta}/2} \Psi_{v}^{i} \approx \bigl(1+i\vec{\sigma}\!\cdot\!\vec{\theta}/2\bigr) \Psi_{v}^{i}$.
Notice that $\Psi_{v}^{i}$ ($i=1,2,3$) is in the rest frame.
Then, we find that $\Psi_{v1/2}$ and $\Psi_{v3/2}^{i}$ change to $\Psi_{v1/2}+\delta\Psi_{v1/2}$ and $\Psi_{v3/2}^{i}+\delta\Psi_{v3/2}^{i}$, respectively, where $\delta\Psi_{v1/2}$ and $\delta\Psi_{v3/2}^{i}$ are given by
\begin{eqnarray}
   \delta\Psi_{v1/2}
=
   \frac{1}{\sqrt{3}} \sigma^{i}\delta\Psi_{v}^{i}
=
    -\frac{i}{6}\theta^{i}\sigma^{i}\Psi_{v1/2}+\frac{i}{2\sqrt{3}}i\varepsilon^{ijk}\theta^{i}\sigma^{j}\Psi_{v3/2}^{k},
\end{eqnarray}
and
\begin{eqnarray}
   \delta\Psi_{v3/2}^{i}
=
   \biggl(\delta^{ij}-\frac{1}{3}\sigma^{i}\sigma^{j}\biggr)\delta\Psi_{v}^{j}
=
- \frac{1}{\sqrt{3}} \frac{i}{2} \theta^{j} \biggl( \frac{4}{3}\delta^{ij} - \frac{2}{3}i\varepsilon^{ijk}\sigma^{k} \biggr) \Psi_{v1/2}
+ \frac{i}{2} \theta^{k} \biggl( \frac{2}{3}\delta^{ij}\sigma^{k} + \frac{1}{3}\delta^{ik}\sigma^{j} - \frac{1}{3} \delta^{jk} \sigma^{i} - \frac{1}{3} i\varepsilon^{ijk} \biggr) \Psi_{v3/2}^{j}.
\nonumber \\
\end{eqnarray}

In terms of the isospin operator $t^{a}$ ($a=1,2,3$) for $\Sigma_{c}$ and $\Sigma_{c}^{\ast}$ baryons, 
the basis used in Eq.~\eqref{eq:t_def} may not be suitable for describing the charged particles such as $\Sigma_{c}^{++}$, $\Sigma_{c}^{+}$, and $\Sigma_{c}^{0}$, because none of $t^{1}$, $t^{2}$, and $t^{3}$ is diagonal.
Instead, it can be useful to introduce the following operator for isospin $\hat{t}^{a}$ ($a=1,2,3$):
\begin{eqnarray}
   \hat{t}^{1}
=
\left(
\begin{array}{ccc}
 0 & \frac{-1}{\sqrt{2}} & 0 \\
 \frac{-1}{\sqrt{2}} & 0 & \frac{1}{\sqrt{2}} \\ 
 0 & \frac{1}{\sqrt{2}} & 0
\end{array}
\right),
\hspace{1em}
   \hat{t}^{2}
=
\left(
\begin{array}{ccc}
 0 & \frac{i}{\sqrt{2}} & 0 \\
 \frac{-i}{\sqrt{2}} & 0 & \frac{-i}{\sqrt{2}} \\ 
 0 & \frac{i}{\sqrt{2}} & 0
\end{array}
\right),
\hspace{1em}
   \hat{t}^{3}
=
\left(
\begin{array}{ccc}
 1 & 0 & 0 \\
 0 & 0 & 0 \\ 
 0 & 0 & -1
\end{array}
\right),
\end{eqnarray}
which are related to $t^{a}$ by the unitary transformation $t^{a}=U\hat{t}^{a}U^{\dag}$ with the unitary matrix
\begin{eqnarray}
   U
=
\left(
\begin{array}{ccc}
 \frac{-i}{\sqrt{2}} & 0 & \frac{-i}{\sqrt{2}} \\
 \frac{1}{\sqrt{2}} & 0 & \frac{-1}{\sqrt{2}}  \\
 0 & -i & 0  
\end{array}
\right).
\end{eqnarray}
We note that the commutation relation holds: $[\hat{t}^{a},\hat{t}^{b}]=i\varepsilon^{abc}\hat{t}^{c}$.
Then, we obtain the new field $\hat{\Psi}_{v1/2}$ and $\hat{\Psi}_{v3/2}^{i}$ expressed by the charge basis:
\begin{eqnarray}
   \hat{\Psi}_{v1/2}
=
\left(
\begin{array}{c}
 \Sigma_{c}^{++} \\
 \Sigma_{c}^{+} \\
 \Sigma_{c}^{0}
\end{array}
\right),
\hspace{1em}
   \hat{\Psi}_{v3/2}^{i}
=
\left(
\begin{array}{c}
 \Sigma_{c}^{\ast++} \\
 \Sigma_{c}^{\ast+} \\
 \Sigma_{c}^{\ast0}
\end{array}
\right),
\end{eqnarray}
which are related to $\Psi_{v1/2}$ and $\Psi_{v3/2}^{i}$ through the unitary transformation $\Psi_{v1/2}=U^{\dag}\hat{\Psi}_{v1/2}$ and $\Psi_{v3/2}^{i}=U^{\dag}\hat{\Psi}_{v3/2}^{i}$.

\bibliography{reference}

\end{document}